\documentclass[bibyear]{aa} 

\usepackage{graphicx, float}
\usepackage{txfonts}
\usepackage{orcidlink}
\usepackage{hyperref}
\RequirePackage[all]{hypcap}
\hypersetup{
    colorlinks=true,
    linkcolor=blue,
    filecolor=blue,      
    urlcolor=blue,
    citecolor=blue
    }
\usepackage{natbib,twoopt}
\bibpunct{(}{)}{;}{a}{}{,} 
\makeatletter
  \newcommandtwoopt{\citeads}[3][][]{\href{http://adsabs.harvard.edu/abs/#3}%
    {\def\hyper@linkstart##1##2{}%
     \let\hyper@linkend\@empty\citealp[#1][#2]{#3}}}
  \newcommandtwoopt{\citepads}[3][][]{\href{http://adsabs.harvard.edu/abs/#3}%
    {\def\hyper@linkstart##1##2{}%
     \let\hyper@linkend\@empty\citep[#1][#2]{#3}}}
  \newcommandtwoopt{\citetads}[3][][]{\href{http://adsabs.harvard.edu/abs/#3}%
    {\def\hyper@linkstart##1##2{}%
     \let\hyper@linkend\@empty\citet[#1][#2]{#3}}}
  \newcommandtwoopt{\citeyearads}[3][][]%
    {\href{http://adsabs.harvard.edu/abs/#3}
    {\def\hyper@linkstart##1##2{}%
     \let\hyper@linkend\@empty\citeyear[#1][#2]{#3}}}
\makeatother
\begin{document}

\title{Effects of physical conditions on the stellar initial mass function: The low-metallicity star-forming region Sh 2-209}

   \author{Marie Zinnkann
          \inst{1}\orcidlink{0009-0000-4037-7118} \and Henriette Wirth  \inst{2}\orcidlink{0000-0003-1258-3162} \and Pavel Kroupa \inst{2,3}\orcidlink{0000-0002-7301-3377}
          }

   \institute{Argelander Institute for Astronomy, University of Bonn,
              Auf dem H\"ugel 71, 53121 Bonn, Germany\\
              \email{s6mazinn@uni-bonn.de}
        \and
        Faculty of Mathematics and Physics, Astronomical Institute, Charles University, V Hole\v{s}ovi\v{c}k\'{a}ch 2, 18000 Praha, Czech Republic
        \email{wirth@sirrah.troja.mff.cuni.cz}
         \and
             Helmholtz Institut f\"ur Strahlen- und Kernphysik, Universit\"at Bonn, Nussallee 14-16, 53115 Bonn, Germany\\
             \email{pkroupa@uni-bonn.de}
             }
   \date{Received 1 August , 2023 / Accepted 26 January 2024}

\authorrunning{M. Zinnkann et al.}
\titlerunning{The physical variation of the stellar initial mass function}
 
  \abstract{Recent work suggested that the variation of the initial mass function (IMF) of stars depends on the physical conditions, notably, the metallicity and gas density. We investigated the properties of two clusters, namely the main cluster (MC) and the subcluster (SC), in the low-metallicity HII region Sh 2-209 (S209) based on recently derived IMFs. We tested three previously published correlations using previous observations: the top-heaviness of the IMF in dependence on metallicity, the half-mass radius, and the most massive star in dependence on the stellar mass of the embedded clusters. For this region, two different galactocentric distances, namely $10.5\,\mathrm{kpc}$ and $18\,\mathrm{kpc}$, were considered, where an age-distance-degeneracy was found for the previously determined IMF to be consistent with other formulated metallicity and density dependent IMFs. The determined half-mass radius $r_\mathrm{h} \approx (0.080 \pm 0.005)\,\mathrm{pc}$  and the embedded cluster density $\rho_\text{ecl} \approx (0.2 \pm 0.1)\,10^6 M_\odot \mathrm{pc}^{-3}$ for the MC with an age of 0.5 Myr in S209 assuming a galactocentric distance of $18\,\mathrm{kpc}$ support the assumption that a low-metallicity environment results in a denser cluster, which leads to a top-heavy IMF. Thus, all three tests are consistent with the previously published correlations. The results for S209 are placed in the context with the IMF determination within the metal-poor cluster in the star-forming region NGC 346 in the Small Magellanic Cloud.}
  
   \keywords{methods: analytical -- binaries: general -- stars: formation -- stars: luminosity function, mass function -- open clusters and associations: individual: Sh 2-209 -- open clusters and associations: individual: NGC 346}

   \maketitle
%
\section{Introduction}
The initial mass function (IMF) gives information about the initial mass distribution of stars in a stellar system (e.g. \citealp{Bastian_IMFvariations}; \citealp{Kroupa_IMF_history}; \citealp{Hopkins_IMF}). Together with previously formulated relations, it describes the mass distribution of a population of stars at birth of the embedded cluster. These previously formulated relations are the $m_\mathrm{max}-M_\mathrm{ecl}$ relation (\citealp{Yan_Mmax_Mecl_relation}), with the maximum stellar mass $m_\mathrm{max}$ and the stellar mass of the embedded cluster $M_\mathrm{ecl}$, and the variation of the IMF with metallicity and density of a molecular cloud core in which the population forms as an embedded star cluster (\citealp{Marks_topheavyIMF}; \citealp{Jerabkova_metallicitydependence}; \citealp{Yan}). Another relation is the $r_\mathrm{h}-M_\mathrm{ecl}$ relation by \cite{Marks}, in which $r_\mathrm{h}$ is the half-mass radius of the embedded cluster at birth, just prior to gas expulsion.
The concepts and relations are explained in more detail in Sect.~\ref{sec:previous_results}.

With this contribution, we test whether the previously determined constraints on the variation of the IMF with metallicity and density of a star-forming gas cloud are consistent with the recently published observations of two young star clusters by \cite{Yasui}. These clusters lie in the low-metallicity star-forming region Sh 2-209, which is located in the outer region of the Milky Way. Furthermore, we compare the observed clusters to the relation between $r_\mathrm{h}$ and  $M_\mathrm{ecl}$, obtained by \cite{Marks}.

The IMFs of solar metallicity regions have been well-studied.
One example is the Taurus region (e.g. \citealp{Luhman_Taurus}; \citealp{Briceno_Taurus}; \citealp{Thies_Taurus}) with a solar metallicity $[\mathrm{Fe}/\mathrm{H}]~=~-0.01\pm0.05$ (\citealp{DOrazi_Taurus}) or NGC 2264 by \cite{Sung_NGC2264_IMF} with $[\mathrm{Fe}/\mathrm{H}] \approx -0.13$ (\citealp{Netopil_NGC2264}).
Another solar metallicity region is the Trapezium cluster in the Orion nebular cluster (ONC) with $[\mathrm{Fe}/\mathrm{H}] = -0.01 \pm 0.04$ (\citealp{DOrazi_Orion_metallicity}), for which \cite{Muench_Trapezium_fullIMF} derived the mass function from B stars to the deuterium-burning limit.
On the other hand, the IMF of an entire low-metallicity region is usually only known over a limited stellar-mass range. 
For example, \cite{Harayama_NGC3603} determined the IMF of the NGC 3603 star cluster within the NGC 3603 HII region in the Milky Way (\citealp{Goss_NGC3603}) within a stellar mass range of 0.4--20 $M_\odot$.
\cite{Kuncarayakti_NGC3603} found it to have a metallicity between $Z=0.004$ and $0.008$.
Because the two-body relaxation time for high-mass stars is similar to the cluster age, \cite{Harayama_NGC3603} suggested that a composition of primordial and dynamical effects leads to the observed top-heaviness of the IMF.
Another example is NGC 346. This star-forming region in the Small Magellanic Cloud (SMC) has a metallicity of $[\mathrm{Fe}/\mathrm{H}] \approx -0.72$ (\citealp{Rochau_NGC346}): the IMF was determined within a stellar mass range of 0.8--60 $M_\odot$ by \cite{Sabbi_NGC346} over a region spanning $\approx40\,\mathrm{pc}$. The authors reported that they did not find any environmental effects on the IMF. 
This holds for the global IMF derived by \cite{Sabbi_NGC346}, who noted its flattening at the centre and attributed it to mass segregation, which must be primordial because NGC 346 is younger than its mass-segregation timescale (see Table 1 in \citealp{Sabbi_NGC346}). This means that more massive stars formed in the centre, as underlined by Fig. 5 in \cite{Sabbi_NGC346}, which supports the idea of variations of the IMF. We discuss NGC 346 in Sect.~\ref{subsec:Discussion_NGC346}.

S209 is the first low-metallicity ($[\mathrm{O}/\mathrm{H}]=-0.5\,\mathrm{dex}$) star-forming region whose IMF was determined over the wide mass range of 0.1--20 $M_\odot$ (\citealp{Yasui}). 
Our approach is to assume the previously constrained variation of the IMF (see Sect.~\ref{subsec:varying_IMF}) and apply it to the two clusters, the main cluster (MC) and the sub-cluster (SC), given the available information on their metallicities. 
With this, we calculate the radii of the two clusters and check them for consistency with the variable IMF and the independently derived relation of $r_\text{h}$ and $M_\text{ecl}$ by \cite{Marks} (see Sect.~\ref{subsec:mass_radius}). 
This places constraints on the distances of both clusters.
We also check for consistency in the relation of $m_\mathrm{max}$ versus the $M_\mathrm{ecl}$ (see Sect.~\ref{subsec:mmax_Mecl}) published by \cite{Yan_Mmax_Mecl_relation}. Section~\ref{sec:previous_results} documents the previously constrained birth relations, Sect.~\ref{sec:Sh209} introduces the observational data, Sect.~\ref{sec:methods} explains the methods, and Sect.~\ref{sec:Results} reports the results. The discussion and conclusion are provided in Sects.~\ref{sec:Discussion} and \ref{sec:conclusion}, respectively.

\section{Previous results on the birth relations}\label{sec:previous_results}
In this section, we give an overview of the  results from previous research on the properties of newly formed embedded clusters that we applied to the data of S209.

\subsection{The canonical initial mass function}\label{subsec:canonical_IMF}
In a stellar system, the initial mass distribution of the stars is of special interest to resolve the formation process and the stellar system's properties. The IMF gives the number of stars with a certain mass. While Salpeter introduced a one-power-law IMF (\citealp{Salpeter_IMF}), \cite{Kroupa_variing_IMF} found that the mass distribution among stars in the solar neighbourhood follows a two-part power-law formulation, the canonical IMF. It is defined as the number of stars in the mass interval $m$ to $m+\mathrm{d}m$, $\mathrm{d} N = \xi(m)\, \mathrm{d}m$, where $\xi (m) = k \cdot k_i m^{-\alpha_i}$, $k$ and $k_i$ are constants, and the exponents are $\alpha_1 = 1.3$ for $0.08 \le m/M_\odot < 0.5$ and $\alpha_2 = 2.3$ for $0.5 \le m/M_\odot < m_\text{max}$, $m$ is the stellar mass, and $m_\mathrm{max}$ is the most massive star in the embedded cluster (Sect.~\ref{subsec:mmax_Mecl}). 

\subsection{The $m_\mathrm{max} - M_\mathrm{ecl}$  relation}\label{subsec:mmax_Mecl}
Based on the IMF being a probability density distribution function (PDF), it might be thought that the IMF of one massive star cluster is similar to that of multiple small star clusters. This contradicts the fact that in a massive embedded star cluster, more massive stars can form, which is not possible in low-mass clusters because the mass of a very massive star would exceed the mass of the small cluster. This concept is documented by the $m_\mathrm{max}-M_\mathrm{ecl}$ relation (\citealp{Weidner_maximumstellarmasss}). A comparison of the results of the two sampling methods, namely the random and optimal sampling introduced by \cite{Kroupa_IMF_history}, shows that the $m_\mathrm{max}-M_\mathrm{ecl}$ relation has a physical origin and is not a statistical effect (\citealp{Yan_Mmax_Mecl_relation}). With this relation, an estimate of $M_\mathrm{ecl}$ is obtained by only having $m_\mathrm{max}$.

\cite{Yan_Mmax_Mecl_relation} quantified the relation between $m_\text{max}$ and $M_\text{ecl}$ (see Figs. 5 and 6 in \citealp{Yan_Mmax_Mecl_relation}).

\subsection{The varying initial mass function}\label{subsec:varying_IMF}
While many Milky Way and Large Magellanic Cloud clusters follow the canoncial IMF (\citealp{Kroupa_IMF}), subsequent data on the mass-to-light ratios (\citealp{Dabringhausen_depending_IMF}) and the occurrence of low-mass X-ray bright binaries (\citealp{Dabringhausen_binaries}) and of the stellar content of low-concentration globular clusters led to the suggestion that the IMF becomes top-heavy at low metallicity and high gas density (\citealp{Marks_topheavyIMF}). 
Star counts in nearby low-metallicity star-forming regions support this result (\citealp{Schneider}; \citealp{Kalari} at the $2\sigma$ confidence level). Assuming this variation of the IMF, \cite{Marks_topheavyIMF}, \cite{Jerabkova_metallicitydependence}, and \cite{Yan} introduced a metallicity-dependent relation between the initial gas density $\rho_\mathrm{gas}$ and the exponent $\alpha_3$.  

Thus, the exponent $\alpha_3$ is related to the initial average cloud core density $\rho_\text{gas}$ and the initial gas metallicity of the cluster $[\mathrm{Z}/\mathrm{H}]$ (\citealp{Jerabkova_metallicitydependence}; \citealp{Yan}),\begin{align}\label{eq:alpha_3} 
\alpha_3 &= 
    \begin{cases}
    2.3, &y < -0.87 \\
    - 0.41y + 1.94, &y > -0.87
    \end{cases} \indent 1 \geq \frac{m}{M_\odot} < m_\mathrm{max},
    \\ 
y &= -0.14 [\mathrm{Z}/\mathrm{H}] + 0.99\log_{10} (\rho_{\text{gas}} / (10^6\text{M}_\odot\text{pc}^{-3})), \label{eq:y} 
\end{align}
with the metal abundance by mass, $[\mathrm{Z}/\mathrm{H}] = [\mathrm{Fe} / \mathrm{H}] + [\alpha / \mathrm{Fe}]$, $[\mathrm{Fe} / \mathrm{H}]$, and $[\alpha / \mathrm{Fe}]$ being the iron and alpha-element abundances, respectively. The typical value of $[\alpha / \mathrm{Fe}]$ in the Milky Way is 0.3 (\citealp{Forbes_metallicity_03}), which is also assumed in this work.
Rearranging Eqs. (\ref{eq:alpha_3}) and (\ref{eq:y}), we obtain

\begin{align}\label{eq:rho_gas}
    &\log_{10} (\rho_{\text{gas}}) = \Big(y + 0.14 [\mathrm{Z}/\mathrm{H}]\Big) \frac{1}{0.99}, \\
    &\text{with}\,\,\, y = -\frac{\alpha_3 - 1.94}{0.41}, \nonumber
\end{align}
where $\rho_{\text{gas}}$ is the initial average density of the cloud core in units of $10^6\mathrm{M}_\odot /\mathrm{pc}^3$, and $\rho_{\text{ecl}}$ is the stellar mass density of the embedded cluster. These two are connected with the relation $\rho_{\text{ecl}} = \epsilon \cdot \rho_\text{gas}$, where the canonical value of $\epsilon = 0.3$ is adopted for the star formation efficiency (\citealp{Lada_Lada_SFE}; \citealp{Andre_SFR}; \citealp{Megeath_SFE}; \citealp{Banerjee_SFE}; \citealp{Wirth}). Furthermore, $\epsilon$ is assumed to be constant throughout the cluster, such that the gas exactly tracks the stars. We note that $\rho_\mathrm{ecl}$ constitutes the mathematically idealised moment of the highest effective density of the embedded cluster at the idealised time when all stars are born instantly, such that the binary population subsequently evolves to the observed populations (\citealp{Marks}). The real physical embedded cluster has a more complex formation history in which the stars form over one to a few crossing times of the contracting molecular cloud core before it expands due to the expulsion of residual gas (e.g. \citealp{Kroupa_formation_starcluster}; \citealp{DellaCroce_massive_cluster}).

For completeness, based on the observation by \cite{Kroupa_IMF} that $\alpha_1$ and $\alpha_2$ appear to also depend on metallicity, \cite{Marks_topheavyIMF} and \cite{Yan} provided an updated formulation (Eq.~\ref{eq:alpha_3}) that includes $\alpha_1([\mathrm{Z}/\mathrm{H}])$ and $\alpha_2([\mathrm{Z}/\mathrm{H}])$.

\subsection{The $r_\mathrm{h} - M_\mathrm{ecl}$ relation}\label{subsec:mass_radius}

Based on a study of binary populations, \cite{Marks} introduced the canonical relation of $r_\text{h}$ and $M_\text{ecl}$ 
\begin{align}\label{eq:r_h}
    \frac{r_\mathrm{h}}{\text{pc}} = 0.10^{+0.07}_{-0.04} \cdot \Bigg(\frac{M_{\text{ecl}}}{\text{M}_\odot}\Bigg)^{0.13 \pm 0.04}.
\end{align}

This is the half-mass radius of the embedded cluster in the most compact configuration, that is, when it is deeply embedded, and prior to any gas expulsion. 
It is required in order for the emanating open star cluster to have the observed population of binary stars, given that star formation provides a binary fraction of almost $100\,\%$ (\citealp{Kroupa_binaryfraction_a, Kroupa_binaryfraction_b}; \citealp{Marks_binarypopulation}).

\section{Sh 2-209}\label{sec:Sh209}
S209 is an HII region located at a Galactic longitude $151^\circ .6062$ and Galactic latitude $-0^\circ.24$ with right ascension R.A. = 4 11 06.7 and declination Dec = +51 09 44 (\citealp{SIMBAD_Database}). 
With data from \cite{Gaia_COllaboration}, \cite{Yasui} determined an astrometric distance of $\approx 2.5\,\text{kpc}$. Assuming the solar galactocentric distance to be $R_\odot = 8.0\,\text{kpc}$, S209 has a galactocentric distance of about 10.5 kpc (\citealp{Yasui}). On the other hand, \cite{Foster_Distance_Sh209} determined a heliocentric distance of $\approx 10.58\,\text{kpc}$ from radial velocity measurements of the Canadian Galactic Plane Survey in the radio regime. This result is consistent with the value determined by \cite{Chini_Distance_S209}, using photometric and spectroscopic data from exciting stars. The galactocentric distance for this case, again assuming $R_\odot = 8.0 \,\text{kpc}$, is therefore approximately 18 kpc.

\begin{figure}[h!]
    \resizebox{\hsize}{!}{\includegraphics{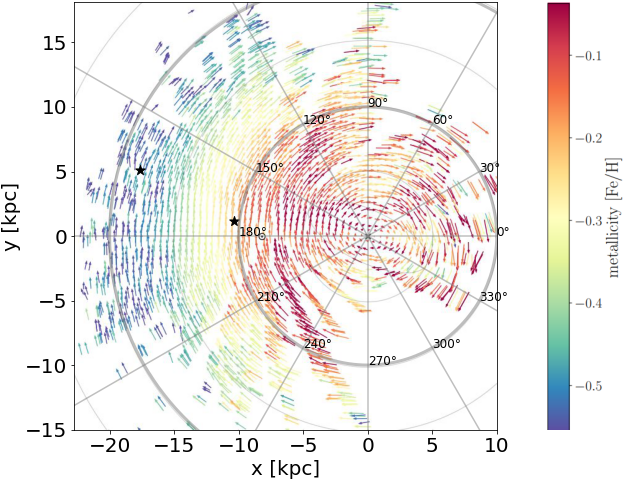}}
    \caption{Metallicity map from \cite{Eilers_Metallicitymap}, where the Galactic centre at (0,0) is indicated by the cross, the Sun is located at (-8.0 kpc,0) and is indicated by the circled dot, and the two possible positions of S209 are indicated by an asterisk. The arrows show the rotation of the stars and are coloured according to their mean self-calibrated metallicity.}
    \label{fig:metallicity_map}
\end{figure}

S209 is a star-forming environment consisting of a MC for which a sample of 1500 objects is identified and a SC with 350 members (\citealp{Yasui}). The observed mass range is 0.1--20 $M_\odot$. Depending on the age and distance, the observed mass range differs. For instance, for younger ages, masses down to $0.02\,M_\odot$ are considered. When the first break mass is lower than the minimum mass of $0.1\,M_\odot$, the first break mass is considered as the minimum mass. Following \cite{Yasui}, we adopt the term 'break mass' as the mass at which the slope of the IMF changes.
The observed mass range was determined from the colour-magnitude diagram. However, some stars that are below this limit but have a large $K$ excess were considered in the derivation of the $K$-band luminosity functions that were used to obtain the IMFs (\citealp{Yasui}). 
To identify the objects in the clusters, near-infrared image data were used, which also show that the cluster is surrounded by residual gas. 
This means that gas expulsion has already occurred. 
Using the metallicity map from \cite{Eilers_Metallicitymap}, we estimated the metallicity of the environment to be $[\text{Fe}/\text{H}]\approx -0.25\,\text{dex}$ for the smaller distance and  $[\text{Fe}/\text{H}]\approx -0.5\,\text{dex}$ for the outer position (see Fig.~\ref{fig:metallicity_map}). Based on previous studies, the result by \cite{Yasui} has an oxygen abundance of $[\mathrm{O}/\mathrm{H}] = -0.5\,\mathrm{dex}$. For a more detailed discussion of the metallicities, we refer to Sect.~\ref{sec:Discussion}.

\section{Methods}\label{sec:methods}
In order to test for the relations we introduced in Sect.~\ref{sec:previous_results}, $M_\mathrm{ecl}$ needs to be calculated.
Ensuring that the IMF is smooth, we determined the constant $k$ using the number of objects $N$ in the star cluster and the equation $N = \int^{m_\text{max}}_{m_\text{min}} \xi(m)\, dm$. 
After constraining the IMF, the mass of the embedded cluster can be computed,

\begin{align}\label{eq:M_ecl}
    \frac{M_\text{ecl}}{M_\odot} ={}&  \int^{m_\text{max}}_{m_\text{min}} m\, \xi(m)\, \mathrm{d}m, \nonumber \\
    ={}& k\cdot \Bigg(\int^{m_1}_{m_\text{min}} m^{\Gamma_1}
    + \frac{m_1^{\Gamma_1}}{m_1^{\Gamma_2}} \int^{m_2}_{m_1} m^{\Gamma_2}
    +\frac{m_1^{\Gamma_1} m_2^{\Gamma_2}}{m_1^{\Gamma_2} m_2^{\Gamma_3}} \int^{m_2}_{m_1} m^{\Gamma_3}\Bigg)\, \mathrm{d}m,
\end{align}
with $\Gamma_i = 1-\alpha_i$.

Here, $m_1$ ($=0.5\,\mathrm{M}_\odot$ for the canonical IMF) is the first break mass, and $m_2$ ($=1\,\mathrm{M}_\odot$ in Eq.~(\ref{eq:alpha_3})) is the second break mass, while $m_\mathrm{min}= 0.08\,\mathrm{M}_\odot$ is approximately the hydrogen-burning limit.

In order to determine $r_\mathrm{h}$, $\rho_\mathrm{ecl}$ needs to be computed. For this, Eqs.~(\ref{eq:y}) and (\ref{eq:rho_gas}) were applied with the known metallicity. Using the relation $\rho_\mathrm{ecl} = \epsilon \cdot \rho_\mathrm{gas}$ with $\epsilon = 0.3$ as explained in Sect.~\ref{subsec:varying_IMF}, $\rho_\mathrm{ecl}$ was obtained.

After determining $M_\mathrm{ecl}$ and $\rho_\mathrm{ecl}$, we calculated the pre-gas-expulsion $r_\mathrm{h}$, assuming spherical symmetry,
\begin{align} \label{eq:r_h_sphere}
    \frac{r_\text{h}}{\text{pc}} = \Bigg(\frac{3\,M_\text{ecl}}{8 \pi \rho_\text{ecl}}\Bigg)^{1/3}.
\end{align}
The results of these calculations are listed in Sect.~\ref{sec:Results}. 

The uncertainties were calculated by Gaussian error propagation ($\Delta f = \sqrt{\sum_i^N 
(\frac{\partial f}{\partial x_i})^2\Delta x_i^2}$, where $\Delta f$ is the uncertainty of the quantity $f$, $x_i$ is a variable, and $\Delta x_i$ is its corresponding error). Because the constant $k$ was calculated via integration and was included in the determination of $M_\mathrm{ecl}$, where a second integration was applied, a large error can be obtained for $M_\mathrm{ecl}$. Furthermore, the errors for the exponents $\Gamma_i$ contribute strongly. This can cause large uncertainties for the $M_\mathrm{ecl}$ values, which lead to large uncertainties for some other results. Additional uncertainties such as extinction due to the interstellar medium or the intracluster medium were not accounted for here because a correction for them was already applied in the analysis by \cite{Yasui}.

After computing $M_\mathrm{ecl}$ and $r_\mathrm{h}$ with Eqs.~(\ref{eq:M_ecl}) and (\ref{eq:r_h_sphere}) based on the $\rho_\mathrm{gas}$ obtained with Eqs.~(\ref{eq:y}) Eq. (\ref{eq:rho_gas}), we compared the results to the  $r_\mathrm{h}-M_\mathrm{ecl}$ relation described in Sect.~\ref{subsec:mass_radius}. This is displayed in Figs.~\ref{fig:M_r_binary_MC} and \ref{fig:M_r_binary_SC} in Sect.~\ref{sec:Discussion}.

\section{Results}\label{sec:Results}
In \cite{Yasui}, the parameters given for the IMF varied for the different galactocentric distances assumed (10.5 kpc or 18 kpc), for the ages from 0.5 to 10 Myr, and for the considered cluster (MC or SC). In order to give an overview, different cases were considered in this paper. In the main part, the parameters of the best-fit IMF are evaluated, and in the Appendix~\ref{sec:apendix_results} the results for ages of 1, 5, and 10 Myr are listed.

\begin{figure}[h!]
    \resizebox{\hsize}{!}{\includegraphics{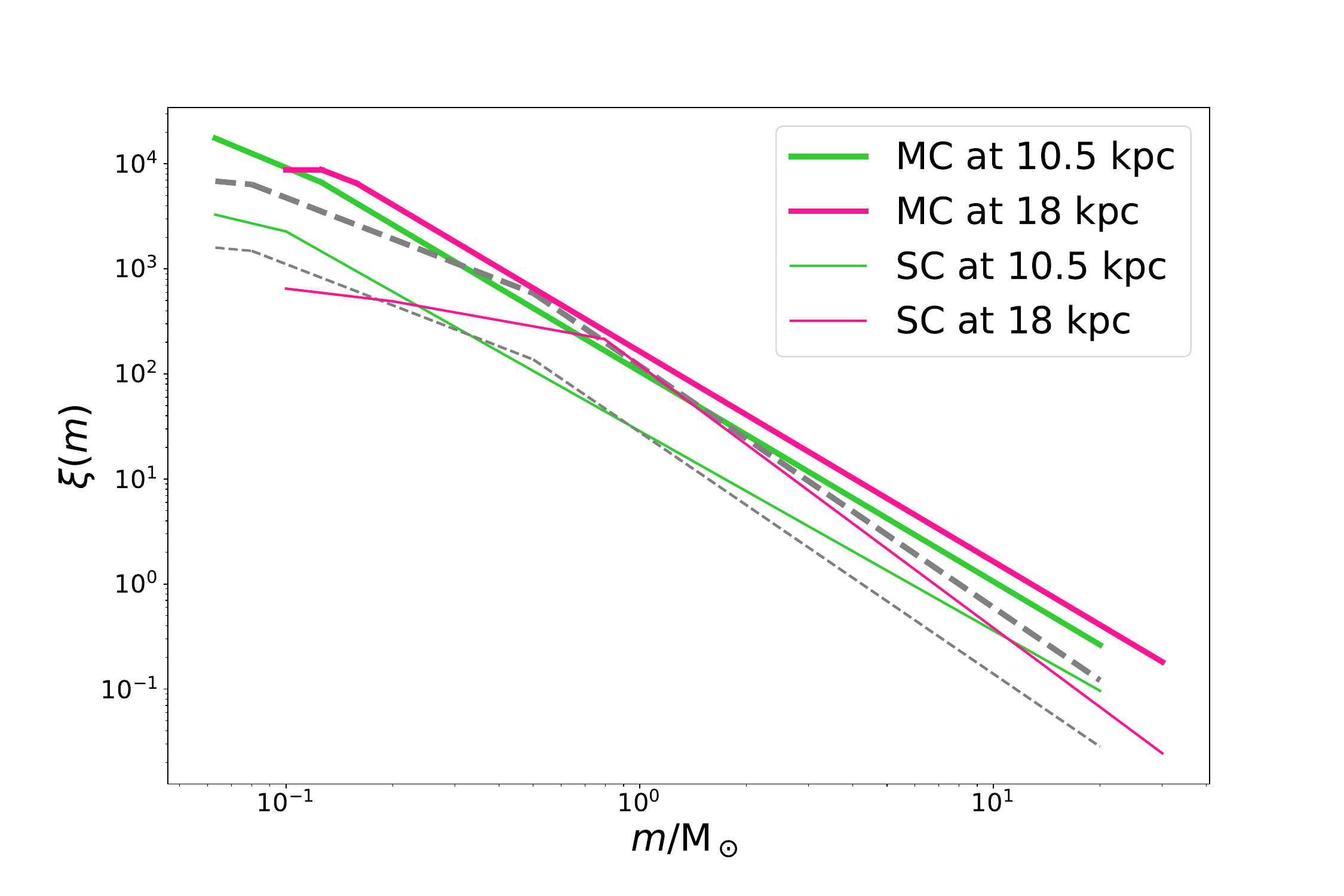}}
    \caption{Reconstructed IMF with values from \cite{Yasui}. The dashed lines represent the canonical IMF (see Sect.~\ref{subsec:canonical_IMF}). The upper line shows the MC, and the lower line shows the SC.}
    \label{fig:IMF}
\end{figure}

Using the best-fit results that \cite{Yasui} obtained, displayed in Table~\ref{tab:main_cluster_values} for the MC and in Table~\ref{tab:sub_cluster_values} for the SC, we plot the IMFs in Fig.~\ref{fig:IMF}. The uncertainties of the parameters were adopted from Table 9 and Sect. 7.1 in \cite{Yasui}.

\begin{figure}[h]
    \resizebox{\hsize}{!}{\includegraphics{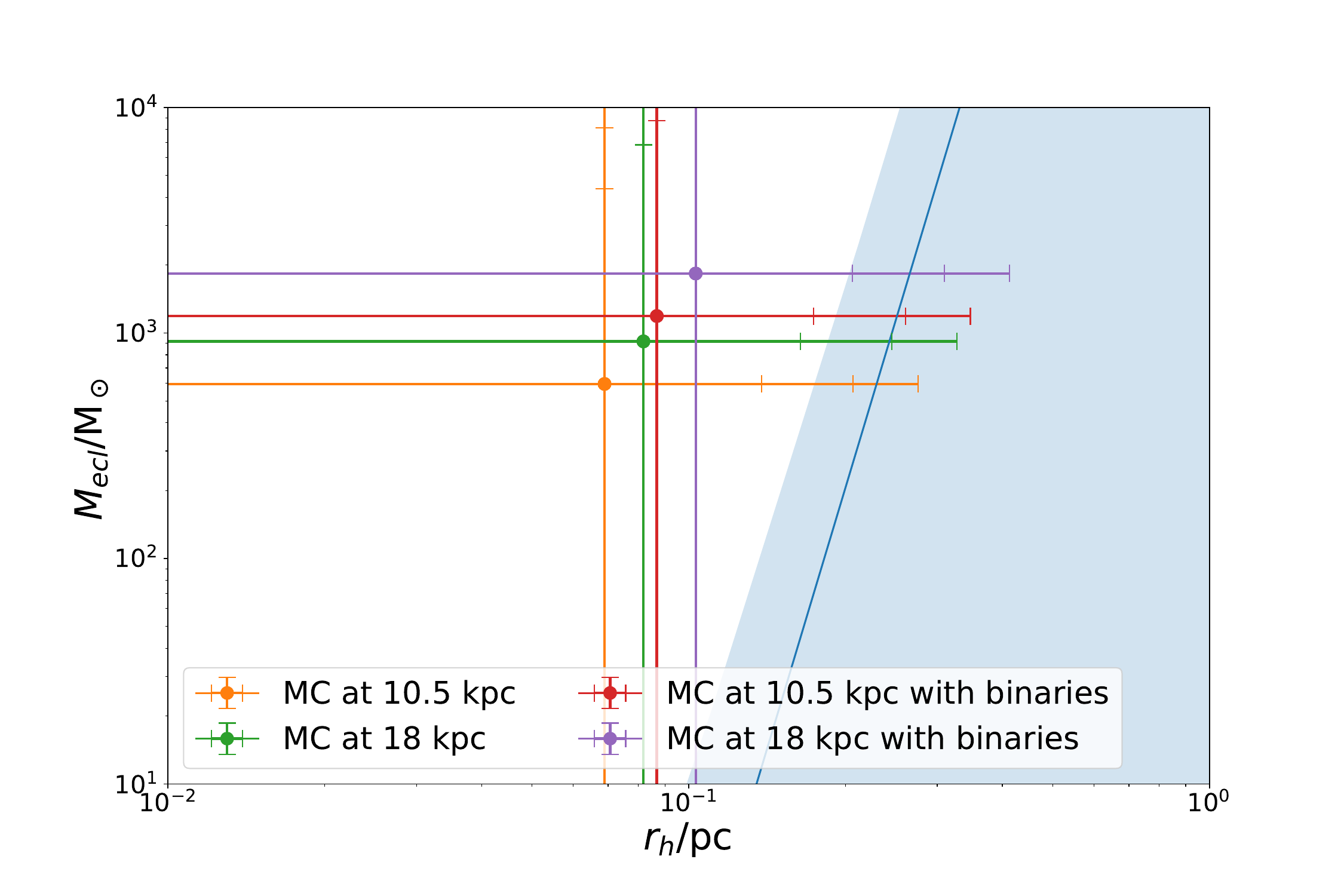}}
    \caption{Canonical relation of $M_\mathrm{ecl}$ and $r_\mathrm{h}$ (Eq.~(\ref{eq:r_h})) shown by the blue line. The data points are the calculated values for the MC in S209. The caps indicate the $1\sigma$ and $2\sigma$ uncertainty boundaries.}
    \label{fig:M_r_binary_MC}
\end{figure}

\begin{table}[h!]
    \caption{Best-fit values for the MC (\citealp{Yasui}).}
    \label{tab:main_cluster_values}
    \centering
    \begin{tabular}{c c c}
    \hline \hline 
     parameter & Distance of 10.5 kpc & Distance of 18 kpc \\
     \hline
     Age/Myr & 5 & 0.5 \\
    $\log_{10} m_1/\text{M}_\odot$ & $-1.2$ & $-0.9$ \\
    $\log_{10} m_2/\text{M}_\odot$  & $-0.9^{+0.1}_{-0.2}$ & $-0.8 \pm 0.1$\\
    $\alpha_1$  & 0 & 0\\
    $\alpha_2$ & $1.4$ & $1.3$\\
    $\alpha_3$ & $2.0 \pm 0.1$ & $2.0 \pm 0.1$ \\
    \end{tabular}
\end{table}

\begin{table}[h!]
    \caption{Best-fit values for the SC (\citealp{Yasui}).}
    \label{tab:sub_cluster_values}
    \centering
    \begin{tabular}{c c c}
    \hline \hline
     parameter & Distance of 10.5 kpc & Distance of 18 kpc \\
     \hline
     Age/Myr & 3 & 1 \\
    $\log_{10} m_1/\text{M}_\odot$ & $-1.2$ & $-0.7$ \\
    $\log_{10} m_2/\text{M}_\odot$  & $-1.0^{+0.1}_{-0.4}$ & $-0.1 \pm 0.2$\\
    $\alpha_1$  & $1.3$ & $0.4$\\
    $\alpha_2$ & $0.8$ & $0.6$\\
    $\alpha_3$ & $1.9^{+0.3}_{-0.2}$ & $2.5 \pm 0.3$ \\
    \end{tabular}
\end{table}

\begin{figure}[H]
    \resizebox{\hsize}{!}{\includegraphics{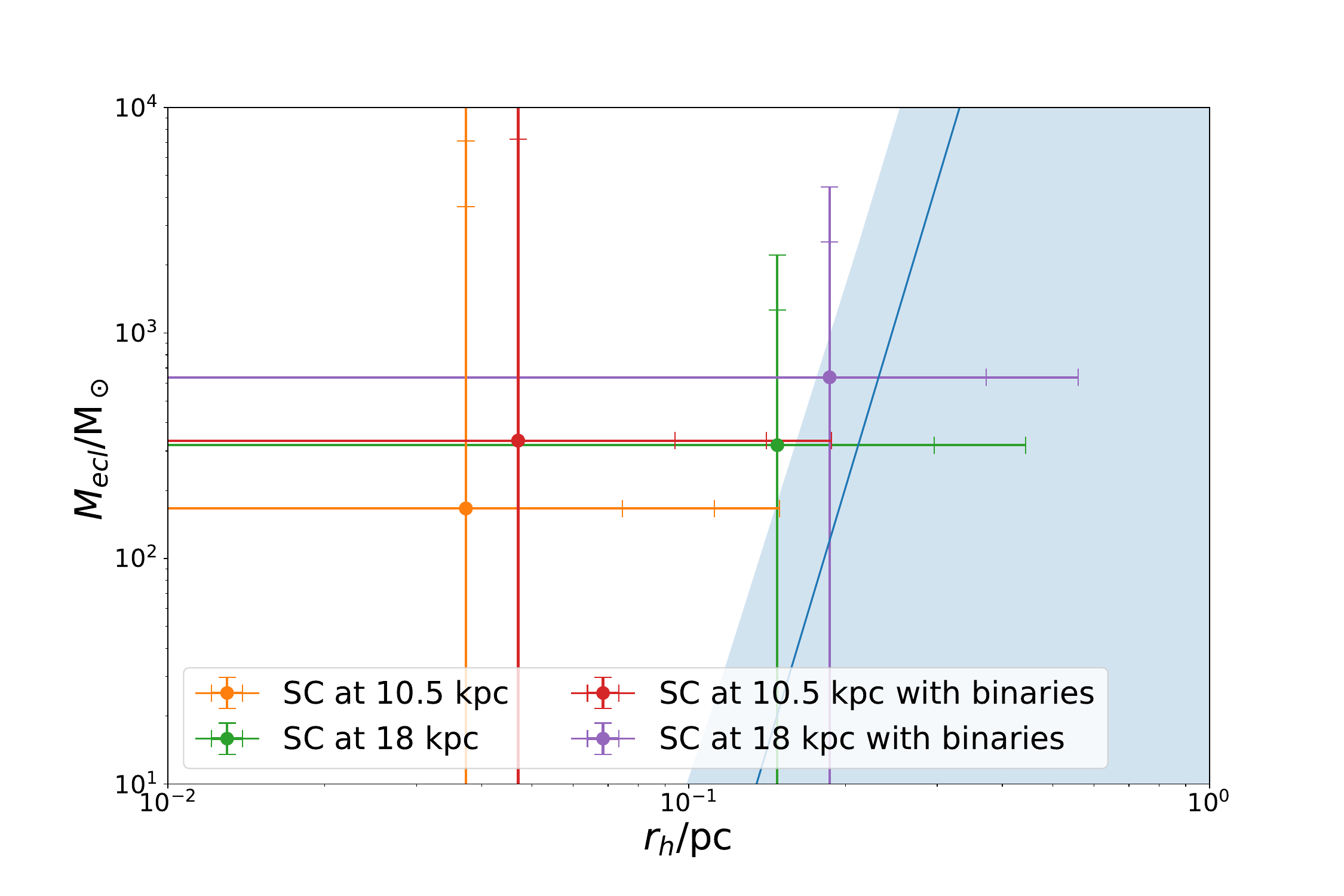}}
    \caption{As Fig.~\ref{fig:M_r_binary_MC}, but for the SC.}
    \label{fig:M_r_binary_SC}
\end{figure}

\noindent We proceeded as follows in order to test whether the observational constraints on $\alpha_3$ by \cite{Yasui} are consistent with the formulation of the variation of the IMF given by Eqs. (\ref{eq:alpha_3}) and (\ref{eq:y}) (Sect.~\ref{subsec:varying_IMF}). By adopting the reported values of $\alpha_3$, Eq. (\ref{eq:rho_gas}) was used to calculate $\rho_\mathrm{gas}$ and Eq. (\ref{eq:M_ecl}) to calculate $M_\mathrm{ecl}$. The application of Eq.~(\ref{eq:r_h_sphere})
yields the value of $r_\mathrm{h}$ that the embedded cluster should have.
The masses of the cluster over these calculated $r_\mathrm{h}$ are displayed in Figs.~\ref{fig:M_r_binary_MC} and \ref{fig:M_r_binary_SC} in Sect.~\ref{sec:Discussion}, where they are compared to the canonical half-mass radius relation (Eq.(~\ref{eq:r_h}) in Sect.~\ref{subsec:mass_radius}).

The resulting densities, radii, and masses are displayed in Table~\ref{tab:main_cluster_results_bestfit} for the MC and in Table~\ref{tab:sub_cluster_results_bestfit} for the SC. The provided results are formal solutions with their $1\sigma$ uncertainties. The $1\sigma$ upper limit would be physically relevant.

Ideally, all stellar masses would be summed from star counts as a constraint on the stellar mass of the cluster, but these data were not available to us. 
\begin{table}[h!]
    \caption{Results for the MC.}
    \label{tab:main_cluster_results_bestfit}
    \centering
    \begin{tabular}{c c c}
    \hline \hline
         parameter & Distance of 10.5 kpc & Distance of 18 kpc  \\
         \hline
         $\rho_\text{gas}/(10^6 \text{M}_\odot \text{pc}^{-3})$ & $0.7 \pm 0.4$ & $0.7 \pm 0.4$\\

         $\rho_\text{ecl}/(10^6 \text{M}_\odot \text{pc}^{-3})$ & $0.2 \pm 0.1$ & $0.2 \pm 0.1$ \\
         
         $M_{\text{ecl}}/\text{M}_\odot$ & $595 \pm 3773$ & $853 \pm 5475$ \\
         
         $r_\text{h}/\text{pc}$ & $0.069 \pm 0.004$ & $0.080 \pm 0.005$ \\
    \end{tabular}
\end{table}

\begin{table}[h!]
    \caption{Results for the SC.}
    \label{tab:sub_cluster_results_bestfit}
    \centering
    \begin{tabular}{c c c}
    \hline \hline
         parameter & Distance of 10.5 kpc & Distance of 18 kpc  \\
         \hline
         $\rho_\text{gas}/(10^6 \text{M}_\odot \text{pc}^{-3})$ & $1 \pm 2$ & $0.04 \pm 0.07$\\

         $\rho_\text{ecl}/(10^6 \text{M}_\odot \text{pc}^{-3})$ & $0.4 \pm 0.7$ & $0.01 \pm 0.02$\\
         
         $M_{\text{ecl}}/\text{M}_\odot$ & $167 \pm 3461$ & $309 \pm 910$ \\
         
         $r_\text{h}/\text{pc}$ & $0.037 \pm 0.004$ & $0.15 \pm 0.02$
    \end{tabular}
\end{table}

\section{Discussion} \label{sec:Discussion}
\subsection{Sh 2-209}
Considering the $m_\mathrm{max}-M_\mathrm{ecl}$ relation that we explained in Sect.~\ref{subsec:mmax_Mecl} and the results for the MC, $m_\text{max}=20\,M_\odot$ (\citealp{Yasui}), we obtain $M_\text{ecl} \approx 800\,M_\odot$, which is compatible with the derived $M_\text{ecl}$ for the best-fit parameters of MC at a galactocentric distance of 18 kpc (see Table~\ref{tab:main_cluster_results_bestfit}). For older ages, it is more consistent with the MC being at a galactocentric distance of 10.5 kpc (see Table~\ref{tab:main_cluster_results_10}). Based on radio emission data, \cite{Richards_cluster_mass} found the ionised gas mass to be greater than $10^3\,M_\odot$. This is more consistent with the values obtained for a galactocentric distance of $10.5\,\mathrm{kpc}$ (the authors even suggest a distance of $4.9\,\mathrm{kpc}$).

With Fig.~\ref{fig:metallicity_map}, we estimated the iron abundance $[\mathrm{Fe}/\mathrm{H}]$ at the two possible positions of S209. \cite{Eilers_Metallicitymap} also provided a map for $[\mathrm{O}/\mathrm{Fe}]$ (see their fig. 6), from which we estimated $[\mathrm{O}/\mathrm{Fe}]\approx 0.03\,\mathrm{dex}$ for the inner and $[\mathrm{O}/\mathrm{Fe}]\approx 0.07\,\mathrm{dex}$ for the outer position. With the relation $[\mathrm{O}/\mathrm{H}] = [\mathrm{O}/\mathrm{Fe}] +[\mathrm{Fe}/\mathrm{H}]$, we obtained an oxygen abundance $[\mathrm{O}/\mathrm{H}]$ of $\approx -0.22\,\mathrm{dex}$ for the smaller and $\approx -0.43\,\mathrm{dex}$ for the larger distance. Because the maps by \cite{Eilers_Metallicitymap} do not account for small-scale variations in the metallicity, the individually observed metallicities may differ from those implied by the \cite{Eilers_Metallicitymap} map. \cite{Yasui} discussed different values for $[\mathrm{O}/\mathrm{H}]$, namely $\approx -0.43\,\mathrm{dex}$ (\citealp{Vilchez_metallicity}), which is consistent with the value derived here. \cite{Rudolph_metallicity} used the data of \cite{Vilchez_metallicity} and derived  $[\mathrm{O}/\mathrm{H}]\approx -0.29\,\mathrm{dex}$, which is $30\%$ lower than the value for the inner position and about $50\%$ higher than the value for the larger distance. Using the work of \cite{Caplan_metallicity}, \cite{Rudolph_metallicity} derived an oxygen abundance of $\approx -0.58\,\mathrm{dex}$ , which is inconsistent with the metallicity determined using the map by \cite{Eilers_Metallicitymap} for the inner position and is about $35\%$ lower than the value for the outer position. 
Finally, \cite{Yasui} suggested $[\mathrm{O}/\mathrm{H}] \approx -0.5\,\mathrm{dex}$, which is consistent with \cite{Eilers_Metallicitymap} for the outer position of S209. This maintains the preference that S209 has a galactocentric distance of 18 kpc, as derived by \cite{Foster_Distance_Sh209} and \cite{Chini_Distance_S209} using spectroscopic and photometric data.

In the determination above, no binary stars are assumed. In the binary-star theorem by \cite{Kroupa_Binary_Theorem}, the majority of star formation results in binary systems. Therefore, a binary fraction of $95\mathrm{\%}$ is adopted in the following, which corresponds to the updated pre-main-sequence eigenevolution model by \cite{Belloni_binarypopulation}. In order to give a maximum correction, we assumed the binaries to be composed of equal-mass stars, but in reality, the mass of the secondaries is somewhat lower than that of the primaries (\citealp{Moe_binaries}; \citealp{Sana_binary}). The value of $\alpha_3$ is not affected by a realistic binary, triple, and quadrupole population (\citealp{Weidner_binary_influence}; \citealp{Kroupa_binary_influence}). Hence, the mass $M_\text{ecl}$ is nearly doubled. Because the density is determined by the metallicty and $\alpha_3$ (see Eqs.~(\ref{eq:alpha_3}) and (\ref{eq:rho_gas})), which means that it is independent of $M_\mathrm{ecl}$, the half-mass radius in Eq. (\ref{eq:r_h_sphere}) increases accordingly.

The canonical relation between $r_\mathrm{h}$ and $M_\mathrm{ecl}$ (Eq.~(\ref{eq:r_h})) and the corresponding results are plotted in Fig.~\ref{fig:M_r_binary_MC} for the MC and in Fig.~\ref{fig:M_r_binary_SC} for the SC. The plots for the other ages are displayed in Appendix~\ref{sec:apendix_results}. In Fig~\ref{fig:M_r_binary_MC}, the value for a galactocentric distance of 18 kpc, which respects the binary-star theorem, lies in the error band and is therefore consistent within $1\sigma$ confidence, while the data points of the galactocentric distance of $10.5\,\mathrm{kpc}$ deviate more from the canonical Eq. (\ref{eq:r_h}), but still lie in the $1\sigma$ regime. 
The deviation of the result for a galactocentric distance of $10.5\,\mathrm{kpc}$ decreases with the binary assumption. 
In the plots in which the other ages are younger than 10 Myr, which are displayed in Appendix~\ref{sec:apendix_results}, the data points for the MC at a galactocentric distance of 18 kpc fit the relation better, while the points for a galactocentric distance of 10.5 kpc do not deviate by more than $2\sigma$. 
This changes for an older age (see Fig.~\ref{fig:M_r_binary_MC_10}), where a galactocentric distance of 10.5 kpc corresponds better to the relation by \cite{Marks}. 

All the half-mass radii are smaller than expected from the canonical Eq. (\ref{eq:r_h}), except for the age of 10 Myr (see Fig.~\ref{fig:M_r_binary_MC_10}). This underlines the possibility that a low-metallicity star-forming region might be denser because the mass is included in a smaller volume than is determined for other very young star clusters of higher metallicity. This would be reminiscent of the smaller radii of low-metallicity stars compared to metal-rich stars of the same mass, suggesting that self-regulation may play an important role for the constitution of an embedded cluster (see also \citealp{Yan_Mmax_Mecl_relation}).

In Fig.~\ref{fig:M_r_binary_SC}, the values for a galactocentric distance of 18 kpc of the SC again agree with the relation by \cite{Marks} within the $1\sigma$ confidence region. The results for the 10.5 kpc distance are only consistent with Eq. (\ref{eq:r_h}) within the $3\sigma$ confidence. Hence, the prediction by \cite{Marks}, that is, Eq. (\ref{eq:r_h}), agrees with the determined values for the MC and SC for both distances, while the results for a galactocentric distance of 18 kpc represent the observational data better. However, given the large uncertainties, no highly significant conclusions can be drawn.

\subsection{NGC 346}\label{subsec:Discussion_NGC346}
NGC 346 is the largest active star-forming region in the Small Magellanic Cloud 
(\citealp{Rickard_NGC346}), for which \cite{Sabbi_NGC346} argued that the IMF is the same as in the Galaxy based on their survey, which spanned $\approx40\,\mathrm{pc}$ around the central embedded cluster. 
Figure 5 in \cite{Sabbi_NGC346} showed, however, that the centremost 4 pc have a top-heavy IMF with $\alpha_3~=~(2.03 \pm 0.14)$, which is
consistent with a star cluster density of $\rho_\mathrm{ecl} \approx (0.25 \pm  0.20) \cdot 10^6 M_\odot \mathrm{pc}^{-3}$, according to Eq.~(\ref{eq:rho_gas}), assuming a
metallicity $[\mathrm{Fe}/\mathrm{H}] \approx -0.72$ (\citealp{Rochau_NGC346}), which we adopted for $[\mathrm{Z}/\mathrm{H}]$ assuming an error of 0.3 to account for  $[\alpha/\mathrm{H}]$. 
With this, the total cluster mass is $M_\mathrm{Tot}\approx 3.9\cdot10^5\,M_\odot$, and with Eq. (\ref{eq:r_h_sphere}), we can give an upper limit on the half-mass radius of the central region of $r_\mathrm{h} < (0.7 \pm 0.2)\,\mathrm{pc}$, which is consistent with the foregoing calculation when the uncertainties caused by  gas expulsion that lead to expansion are included. 
At larger distances between 4 and 9 pc, \cite{Sabbi_NGC346} reported a canonical IMF ($\alpha_3 \approx 2.36\pm0.09$), which is likely a mixture of on-site lower-mass embedded clusters and ejected massive stars from the starburst cluster (\citealp{Oh_starejection}; \citealp{Oh_Ostarejection}). At larger distances, $\alpha_3 \approx (2.43 \pm 0.09)$ and for distances larger than 14 pc, $\alpha_3 \approx (3.08 \pm 0.14)$, which reflects a population of low-mass embedded clusters lacking massive stars (\citealp{Weidner_maximumstellarmasss}; \citealp{Yan_Mmax_Mecl_relation}). Although the ejected massive stars induce a steeper slope of the MF in the inner part, mass segregation is more present in this case, which leads to the flatter slope in the central region.
The centre of this whole star-forming system on a scale of $\approx 40\,\mathrm{pc}$ across is reminiscent of the Orion-South star-forming cloud where only the ONC formed massive stars, while much of the southern part of the cloud formed many low-mass embedded clusters lacking massive stars (\citealp{Hsu_ONC}).
In contrast to the assertion by \cite{Sabbi_NGC346}, NGC 346 thus appears to be consistent with the variation of the IMF given by Eqs.~(\ref{eq:alpha_3}) and (\ref{eq:y}), but more work is needed to quantify details. For example, it will be necessary to quantify whether many low-mass embedded clusters formed with IMFs lacking massive stars compared to the centred starburst cluster at larger distances from the centre of NGC 346. 

\section{Conclusions}\label{sec:conclusion}
We analysed two clusters of the star-forming region S209 with respect to their initial conditions. In particular, the low metallicity was taken into account, which allowed us to determine $\rho_\text{ecl}$ (Eq.~(\ref{eq:rho_gas})), by using the determination of the IMF power-law index by \cite{Yasui}. 
We constrained $r_\text{h}$ and compared it to the independently developed canonical relation of $r_\text{h}$ and $M_\text{ecl}$ (Eq.~(\ref{eq:r_h})).

Thus, the IMF data from \cite{Yasui} for the low-metallicity region S209 result in a top-heavy IMF. The determined value of $M_\mathrm{ecl}$ for ages of about 0.5 Myr for the MC at a galactocentric distance of 18 kpc is most consistent with the value $M_\mathrm{ecl}\approx 800\,M_\odot$ deduced with the relation of $m_\mathrm{max}$ and $M_\mathrm{ecl}$ derived by  \cite{Yan_Mmax_Mecl_relation}. This changes for older ages, where the resulting  mass, $M_\mathrm{ecl}$, for the smaller galactocentric distance of $10.5\,\mathrm{kpc}$ is more consistent with the $m_\mathrm{max}-M_\mathrm{ecl}$ relation. 

The comparison of the determined metallicity with the map by \cite{Eilers_Metallicitymap} for the outer position and the independently derived metallicities by \cite{Yasui} and  \cite{Vilchez_metallicity} indicate a preferred galactocentric distance of 18 kpc. The value for the inner position (10.5 kpc) is not consistent with the metallicity of S209.

The calculated value for the $r_\text{h}$ of the MC at a galactocentric distance of 18 kpc is consistent with the canonical $r_\mathrm{h}-M_\mathrm{ecl}$ relation for young ages, namely being within the $1\sigma$ confidence region. The binary theorem introduced by \cite{Kroupa_Binary_Theorem} tightens these results. The values of $r_\mathrm{h}$ for the MC and SC at a  galactocentric distance of 10.5 kpc we found are only consistent within $2\sigma$ or $3\sigma$ confidence for young ages. On the other hand, they replicate the $m_\mathrm{max}-M_\mathrm{ecl}$ relation (\citealp{Yan}) better for an age of 10 Myr. This confirms the previously mentioned age-distance degeneracy.

In addition, we showed that the variation of the IMF (Eq.~(\ref{eq:y})) is consistent with NGC 346, which shows a top-heavy IMF in the innermost region and a lack of massive stars with increasing distance to the centre. This is reminiscent of the ONC. 

In conclusion, S209 is consistent with a low-metallicity region resulting in a denser star-forming region with a top-heavy IMF, as quantified by \cite{Marks_topheavyIMF} and \cite{Yan}. The findings for the MC with a galactocentric distance of $18\,\mathrm{kpc}$ are consistent with the relation of $m_\mathrm{max}$ and $M_\mathrm{ecl}$ by \cite{Yan_Mmax_Mecl_relation} for young ages, while a galactocentric distance of $10.5\,\mathrm{kpc}$ is more consistent for an older age. When the binary-star theorem by \cite{Kroupa_Binary_Theorem} is included, the values determined for $M_\mathrm{ecl}$ and $r_\mathrm{h}$ in this paper are consistent with the $r_\mathrm{h} - M_\mathrm{ecl}$ relation derived by \cite{Marks}. To quantify this result further, more low-metallicity environments have to be examined in the future, and the heliocentric distance measurement of S209 needs to be improved.

\begin{acknowledgements}
We acknowledge support through the DAAD-Eastern-Europe Exchange grant at Bonn University and corresponding support from Charles University.
\end{acknowledgements}

%
%
%
\bibliographystyle{aa}
\bibliography{References}

\begin{thebibliography}{60}
\expandafter\ifx\csname natexlab\endcsname\relax\def\natexlab#1{#1}\fi

\bibitem[{{Andr{\'e}} {et~al.}(2014){Andr{\'e}}, {Di Francesco}, {Ward-Thompson}, {Inutsuka}, {Pudritz}, \& {Pineda}}]{Andre_SFR}
{Andr{\'e}}, P., {Di Francesco}, J., {Ward-Thompson}, D., {et~al.} 2014, in Protostars and Planets VI, ed. H.~{Beuther}, R.~S. {Klessen}, C.~P. {Dullemond}, \& T.~{Henning}, 27--51

\bibitem[{{Banerjee} \& {Kroupa}(2018)}]{Banerjee_SFE}
{Banerjee}, S. \& {Kroupa}, P. 2018, in Astrophysics and Space Science Library, Vol. 424, The Birth of Star Clusters, ed. S.~{Stahler}, 143

\bibitem[{{Bastian} {et~al.}(2010){Bastian}, {Covey}, \& {Meyer}}]{Bastian_IMFvariations}
{Bastian}, N., {Covey}, K.~R., \& {Meyer}, M.~R. 2010, \araa, 48, 339

\bibitem[{{Belloni} {et~al.}(2017){Belloni}, {Askar}, {Giersz}, {Kroupa}, \& {Rocha-Pinto}}]{Belloni_binarypopulation}
{Belloni}, D., {Askar}, A., {Giersz}, M., {Kroupa}, P., \& {Rocha-Pinto}, H.~J. 2017, \mnras, 471, 2812

\bibitem[{{Brice{\~n}o} {et~al.}(2002){Brice{\~n}o}, {Luhman}, {Hartmann}, {Stauffer}, \& {Kirkpatrick}}]{Briceno_Taurus}
{Brice{\~n}o}, C., {Luhman}, K.~L., {Hartmann}, L., {Stauffer}, J.~R., \& {Kirkpatrick}, J.~D. 2002, \apj, 580, 317

\bibitem[{{Caplan} {et~al.}(2000){Caplan}, {Deharveng}, {Pe{\~n}a}, {Costero}, \& {Blondel}}]{Caplan_metallicity}
{Caplan}, J., {Deharveng}, L., {Pe{\~n}a}, M., {Costero}, R., \& {Blondel}, C. 2000, \mnras, 311, 317

\bibitem[{{Chini} \& {Wink}(1984)}]{Chini_Distance_S209}
{Chini}, R. \& {Wink}, J.~E. 1984, \aap, 139, L5

\bibitem[{{Dabringhausen} {et~al.}(2009){Dabringhausen}, {Kroupa}, \& {Baumgardt}}]{Dabringhausen_depending_IMF}
{Dabringhausen}, J., {Kroupa}, P., \& {Baumgardt}, H. 2009, \mnras, 394, 1529

\bibitem[{{Dabringhausen} {et~al.}(2012){Dabringhausen}, {Kroupa}, {Pflamm-Altenburg}, \& {Mieske}}]{Dabringhausen_binaries}
{Dabringhausen}, J., {Kroupa}, P., {Pflamm-Altenburg}, J., \& {Mieske}, S. 2012, \apj, 747, 72

\bibitem[{{Della Croce} {et~al.}(2023){Della Croce}, {Dalessandro}, {Livernois}, {Vesperini}, {Fanelli}, {Origlia}, {Bellazzini}, {Oliva}, {Sanna}, \& {Varri}}]{DellaCroce_massive_cluster}
{Della Croce}, A., {Dalessandro}, E., {Livernois}, A., {et~al.} 2023, \aap, 674, A93

\bibitem[{{D'Orazi} {et~al.}(2011){D'Orazi}, {Biazzo}, \& {Randich}}]{DOrazi_Taurus}
{D'Orazi}, V., {Biazzo}, K., \& {Randich}, S. 2011, \aap, 526, A103

\bibitem[{{D'Orazi} {et~al.}(2009){D'Orazi}, {Randich}, {Flaccomio}, {Palla}, {Sacco}, \& {Pallavicini}}]{DOrazi_Orion_metallicity}
{D'Orazi}, V., {Randich}, S., {Flaccomio}, E., {et~al.} 2009, \aap, 501, 973

\bibitem[{{Eilers} {et~al.}(2022){Eilers}, {Hogg}, {Rix}, {Ness}, {Price-Whelan}, {M{\'e}sz{\'a}ros}, \& {Nitschelm}}]{Eilers_Metallicitymap}
{Eilers}, A.-C., {Hogg}, D.~W., {Rix}, H.-W., {et~al.} 2022, \apj, 928, 23

\bibitem[{{Forbes} {et~al.}(2011){Forbes}, {Spitler}, {Strader}, {Romanowsky}, {Brodie}, \& {Foster}}]{Forbes_metallicity_03}
{Forbes}, D.~A., {Spitler}, L.~R., {Strader}, J., {et~al.} 2011, \mnras, 413, 2943

\bibitem[{{Foster} \& {Brunt}(2015)}]{Foster_Distance_Sh209}
{Foster}, T. \& {Brunt}, C.~M. 2015, \aj, 150, 147

\bibitem[{{Gaia Collaboration} {et~al.}(2021){Gaia Collaboration}, {Brown}, {Vallenari}, {Prusti}, {de Bruijne}, {Babusiaux}, {Biermann}, {Creevey}, \& {Evans}}]{Gaia_COllaboration}
{Gaia Collaboration}, {Brown}, A.~G.~A., {Vallenari}, A., {et~al.} 2021, \aap, 649, A1

\bibitem[{{Goss} \& {Radhakrishnan}(1969)}]{Goss_NGC3603}
{Goss}, W.~M. \& {Radhakrishnan}, V. 1969, \aplett, 4, 199

\bibitem[{{Harayama} {et~al.}(2008){Harayama}, {Eisenhauer}, \& {Martins}}]{Harayama_NGC3603}
{Harayama}, Y., {Eisenhauer}, F., \& {Martins}, F. 2008, \apj, 675, 1319

\bibitem[{{Hopkins}(2018)}]{Hopkins_IMF}
{Hopkins}, A.~M. 2018, \pasa, 35, e039

\bibitem[{{Hsu} {et~al.}(2012){Hsu}, {Hartmann}, {Allen}, {Hern{\'a}ndez}, {Megeath}, {Mosby}, {Tobin}, \& {Espaillat}}]{Hsu_ONC}
{Hsu}, W.-H., {Hartmann}, L., {Allen}, L., {et~al.} 2012, \apj, 752, 59

\bibitem[{{Je{\v{r}}{\'a}bkov{\'a}} {et~al.}(2018){Je{\v{r}}{\'a}bkov{\'a}}, {Hasani Zonoozi}, {Kroupa}, {Beccari}, {Yan}, {Vazdekis}, \& {Zhang}}]{Jerabkova_metallicitydependence}
{Je{\v{r}}{\'a}bkov{\'a}}, T., {Hasani Zonoozi}, A., {Kroupa}, P., {et~al.} 2018, \aap, 620, A39

\bibitem[{Kalari {et~al.}(2018)Kalari, Carraro, Evans, \& Rubio}]{Kalari}
Kalari, V.~M., Carraro, G., Evans, C.~J., \& Rubio, M. 2018, The Astrophysical Journal, 857, 132

\bibitem[{{Kroupa}(1995{\natexlab{a}})}]{Kroupa_binaryfraction_a}
{Kroupa}, P. 1995{\natexlab{a}}, \mnras, 277, 1491

\bibitem[{{Kroupa}(1995{\natexlab{b}})}]{Kroupa_binaryfraction_b}
{Kroupa}, P. 1995{\natexlab{b}}, \mnras, 277, 1507

\bibitem[{{Kroupa}(2001)}]{Kroupa_variing_IMF}
{Kroupa}, P. 2001, \mnras, 322, 231

\bibitem[{{Kroupa}(2002)}]{Kroupa_IMF}
{Kroupa}, P. 2002, Astronomical Society of the Pacific Conference Series, 285, 86

\bibitem[{{Kroupa}(2008)}]{Kroupa_Binary_Theorem}
{Kroupa}, P. 2008, in The Cambridge N-Body Lectures, ed. S.~J. {Aarseth}, C.~A. {Tout}, \& R.~A. {Mardling}, Vol. 760, 181

\bibitem[{{Kroupa} {et~al.}(2001){Kroupa}, {Aarseth}, \& {Hurley}}]{Kroupa_formation_starcluster}
{Kroupa}, P., {Aarseth}, S., \& {Hurley}, J. 2001, \mnras, 321, 699

\bibitem[{{Kroupa} \& {Jerabkova}(2018)}]{Kroupa_binary_influence}
{Kroupa}, P. \& {Jerabkova}, T. 2018, arXiv e-prints, arXiv:1806.10605

\bibitem[{{Kroupa} {et~al.}(2013){Kroupa}, {Weidner}, {Pflamm-Altenburg}, {Thies}, {Dabringhausen}, {Marks}, \& {Maschberger}}]{Kroupa_IMF_history}
{Kroupa}, P., {Weidner}, C., {Pflamm-Altenburg}, J., {et~al.} 2013, in Planets, Stars and Stellar Systems. Volume 5: Galactic Structure and Stellar Populations, ed. T.~D. {Oswalt} \& G.~{Gilmore}, Vol.~5, 115

\bibitem[{{Kuncarayakti} {et~al.}(2016){Kuncarayakti}, {Galbany}, {Anderson}, {Kr{\"u}hler}, \& {Hamuy}}]{Kuncarayakti_NGC3603}
{Kuncarayakti}, H., {Galbany}, L., {Anderson}, J.~P., {Kr{\"u}hler}, T., \& {Hamuy}, M. 2016, \aap, 593, A78

\bibitem[{{Lada} \& {Lada}(2003)}]{Lada_Lada_SFE}
{Lada}, C.~J. \& {Lada}, E.~A. 2003, \araa, 41, 57

\bibitem[{{Luhman}(2000)}]{Luhman_Taurus}
{Luhman}, K.~L. 2000, \apj, 544, 1044

\bibitem[{{Marks} \& {Kroupa}(2011)}]{Marks_binarypopulation}
{Marks}, M. \& {Kroupa}, P. 2011, \mnras, 417, 1702

\bibitem[{{Marks} \& {Kroupa}(2012)}]{Marks}
{Marks}, M. \& {Kroupa}, P. 2012, \aap, 543, A8

\bibitem[{{Marks} {et~al.}(2012){Marks}, {Kroupa}, {Dabringhausen}, \& {Pawlowski}}]{Marks_topheavyIMF}
{Marks}, M., {Kroupa}, P., {Dabringhausen}, J., \& {Pawlowski}, M.~S. 2012, \mnras, 422, 2246

\bibitem[{{Megeath} {et~al.}(2016){Megeath}, {Gutermuth}, {Muzerolle}, {Kryukova}, {Hora}, {Allen}, {Flaherty}, {Hartmann}, {Myers}, {Pipher}, {Stauffer}, {Young}, \& {Fazio}}]{Megeath_SFE}
{Megeath}, S.~T., {Gutermuth}, R., {Muzerolle}, J., {et~al.} 2016, \aj, 151, 5

\bibitem[{{Moe} \& {Di Stefano}(2013)}]{Moe_binaries}
{Moe}, M. \& {Di Stefano}, R. 2013, \apj, 778, 95

\bibitem[{{Muench} {et~al.}(2002){Muench}, {Lada}, {Lada}, \& {Alves}}]{Muench_Trapezium_fullIMF}
{Muench}, A.~A., {Lada}, E.~A., {Lada}, C.~J., \& {Alves}, J. 2002, \apj, 573, 366

\bibitem[{{Netopil} {et~al.}(2016){Netopil}, {Paunzen}, {Heiter}, \& {Soubiran}}]{Netopil_NGC2264}
{Netopil}, M., {Paunzen}, E., {Heiter}, U., \& {Soubiran}, C. 2016, \aap, 585, A150

\bibitem[{{Oh} \& {Kroupa}(2016)}]{Oh_starejection}
{Oh}, S. \& {Kroupa}, P. 2016, \aap, 590, A107

\bibitem[{{Oh} {et~al.}(2015){Oh}, {Kroupa}, \& {Pflamm-Altenburg}}]{Oh_Ostarejection}
{Oh}, S., {Kroupa}, P., \& {Pflamm-Altenburg}, J. 2015, \apj, 805, 92

\bibitem[{{Richards} {et~al.}(2012){Richards}, {Lang}, {Trombley}, \& {Figer}}]{Richards_cluster_mass}
{Richards}, E.~E., {Lang}, C.~C., {Trombley}, C., \& {Figer}, D.~F. 2012, \aj, 144, 89

\bibitem[{{Rickard} {et~al.}(2022){Rickard}, {Hainich}, {Hamann}, {Oskinova}, {Prinja}, {Ramachandran}, {Pauli}, {Todt}, {Sander}, {Shenar}, {Chu}, \& {Gallagher}}]{Rickard_NGC346}
{Rickard}, M.~J., {Hainich}, R., {Hamann}, W.~R., {et~al.} 2022, \aap, 666, A189

\bibitem[{{Rochau} {et~al.}(2007){Rochau}, {Gouliermis}, {Brandner}, {Dolphin}, \& {Henning}}]{Rochau_NGC346}
{Rochau}, B., {Gouliermis}, D.~A., {Brandner}, W., {Dolphin}, A.~E., \& {Henning}, T. 2007, \apj, 664, 322

\bibitem[{{Rudolph} {et~al.}(2006){Rudolph}, {Fich}, {Bell}, {Norsen}, {Simpson}, {Haas}, \& {Erickson}}]{Rudolph_metallicity}
{Rudolph}, A.~L., {Fich}, M., {Bell}, G.~R., {et~al.} 2006, \apjs, 162, 346

\bibitem[{{Sabbi} {et~al.}(2008){Sabbi}, {Sirianni}, {Nota}, {Tosi}, {Gallagher}, {Smith}, {Angeretti}, {Meixner}, {Oey}, {Walterbos}, \& {Pasquali}}]{Sabbi_NGC346}
{Sabbi}, E., {Sirianni}, M., {Nota}, A., {et~al.} 2008, \aj, 135, 173

\bibitem[{{Salpeter}(1955)}]{Salpeter_IMF}
{Salpeter}, E.~E. 1955, \apj, 121, 161

\bibitem[{{Sana} {et~al.}(2012){Sana}, {de Mink}, {de Koter}, {Langer}, {Evans}, {Gieles}, {Gosset}, {Izzard}, {Le Bouquin}, \& {Schneider}}]{Sana_binary}
{Sana}, H., {de Mink}, S.~E., {de Koter}, A., {et~al.} 2012, Science, 337, 444

\bibitem[{{Schneider} {et~al.}(2018){Schneider}, {Sana}, {Evans}, {Bestenlehner}, {Castro}, {Fossati}, {Gr{\"a}fener}, {Langer}, {Ram{\'\i}rez-Agudelo}, {Sab{\'\i}n-Sanjuli{\'a}n}, {Sim{\'o}n-D{\'\i}az}, {Tramper}, {Crowther}, {de Koter}, {de Mink}, {Dufton}, {Garcia}, {Gieles}, {H{\'e}nault-Brunet}, {Herrero}, {Izzard}, {Kalari}, {Lennon}, {Ma{\'\i}z Apell{\'a}niz}, {Markova}, {Najarro}, {Podsiadlowski}, {Puls}, {Taylor}, {van Loon}, {Vink}, \& {Norman}}]{Schneider}
{Schneider}, F.~R.~N., {Sana}, H., {Evans}, C.~J., {et~al.} 2018, Science, 359, 69

\bibitem[{{Sung} \& {Bessell}(2010)}]{Sung_NGC2264_IMF}
{Sung}, H. \& {Bessell}, M.~S. 2010, \aj, 140, 2070

\bibitem[{{Thies} \& {Kroupa}(2008)}]{Thies_Taurus}
{Thies}, I. \& {Kroupa}, P. 2008, \mnras, 390, 1200

\bibitem[{{Vilchez} \& {Esteban}(1996)}]{Vilchez_metallicity}
{Vilchez}, J.~M. \& {Esteban}, C. 1996, \mnras, 280, 720

\bibitem[{{Weidner} \& {Kroupa}(2006)}]{Weidner_maximumstellarmasss}
{Weidner}, C. \& {Kroupa}, P. 2006, \mnras, 365, 1333

\bibitem[{{Weidner} {et~al.}(2009){Weidner}, {Kroupa}, \& {Maschberger}}]{Weidner_binary_influence}
{Weidner}, C., {Kroupa}, P., \& {Maschberger}, T. 2009, \mnras, 393, 663

\bibitem[{{Wenger} {et~al.}(2000){Wenger}, {Ochsenbein}, {Egret}, {Dubois}, {Bonnarel}, {Borde}, {Genova}, {Jasniewicz}, {Lalo{\"e}}, {Lesteven}, \& {Monier}}]{SIMBAD_Database}
{Wenger}, M., {Ochsenbein}, F., {Egret}, D., {et~al.} 2000, \aaps, 143, 9

\bibitem[{{Wirth} {et~al.}(2022){Wirth}, {Kroupa}, {Haas}, {Jerabkova}, {Yan}, \& {{\v{S}}ubr}}]{Wirth}
{Wirth}, H., {Kroupa}, P., {Haas}, J., {et~al.} 2022, \mnras, 516, 3342

\bibitem[{{Yan} {et~al.}(2021){Yan}, {Je{\v{r}}{\'a}bkov{\'a}}, \& {Kroupa}}]{Yan}
{Yan}, Z., {Je{\v{r}}{\'a}bkov{\'a}}, T., \& {Kroupa}, P. 2021, \aap, 655, A19

\bibitem[{{Yan} {et~al.}(2023){Yan}, {Je{\v{r}}{\'a}bkov{\'a}}, \& {Kroupa}}]{Yan_Mmax_Mecl_relation}
{Yan}, Z., {Je{\v{r}}{\'a}bkov{\'a}}, T., \& {Kroupa}, P. 2023, \aap, 670, A151

\bibitem[{{Yasui} {et~al.}(2023){Yasui}, {Kobayashi}, {Saito}, {Izumi}, \& {Ikeda}}]{Yasui}
{Yasui}, C., {Kobayashi}, N., {Saito}, M., {Izumi}, N., \& {Ikeda}, Y. 2023, \apj, 943, 137

\end{thebibliography}

\begin{appendix}
\section{Data and results for different ages}\label{sec:apendix_results}

\subsection{Age of 1 Myr}

\begin{table}[h!]
    \caption{Best-fit values for the MC with an age of 1 Myr (\citealp{Yasui}).}
    \label{tab:main_cluster_values_1}
    \centering
    \begin{tabular}{c c c}
    \hline \hline 
     parameter & Distance of 2.5 kpc & Distance of 10 kpc \\
     \hline
    $\log_{10} m_1/\text{M}_\odot$ & $-1.5$ & $-0.8$ \\
    $\log_{10} m_2/\text{M}_\odot$  & $-1.4 \pm 0.1$ & $-0.2 \pm 0.1$\\
    $\alpha_1$  & 0.7 & 1.2\\
    $\alpha_2$ & $1.1$ & 0.9\\
    $\alpha_3$ & $2.3 \pm 0.1$ & $2.2 \pm 0.1$ \\
    \end{tabular}
\end{table}

\begin{table}[h!]
    \caption{Best-fit values for the SC with an age of 1 Myr (\citealp{Yasui}).}
    \label{tab:sub_cluster_values_1}
    \centering
    \begin{tabular}{c c c}
    \hline \hline 
     parameter & Distance of 2.5 kpc & Distance of 10 kpc \\
     \hline
    $\log_{10} m_1/\text{M}_\odot$ & $-1.5$ & $-0.7$ \\
    $\log_{10} m_2/\text{M}_\odot$  & $-1.2 \pm 0.2$ & $-0.1 \pm 0.2$\\
    $\alpha_1$  & $1.1$ & $0.4$\\
    $\alpha_2$ & $0.7$ & 0.6\\
    $\alpha_3$ & $2.6 \pm 0.3$ & $2.5 \pm 0.3$ \\
    \end{tabular}
\end{table}

\begin{figure}[h!]
    \resizebox{\hsize}{!}{\includegraphics{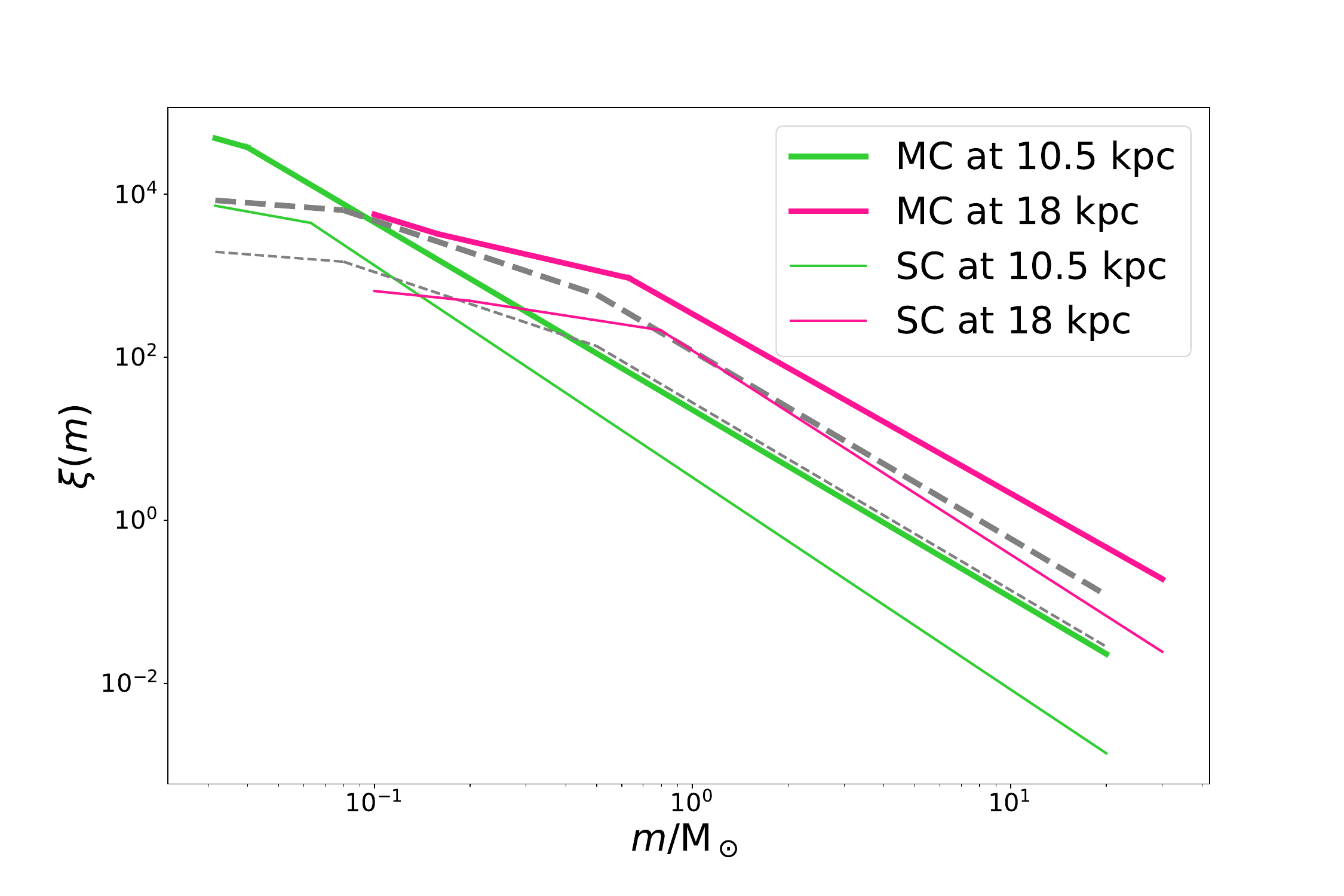}}
    \caption{As Fig.~\ref{fig:IMF}, but for an age of 1 Myr.}
    \label{fig:IMF_1}
\end{figure}

\begin{table}[h!]
    \caption{Results for the MC with an age of $1\,\mathrm{Myr}$.}
    \label{tab:main_cluster_results_1}
    \centering
    \begin{tabular}{c c c}
    \hline \hline
         parameter & Distance of 2.5 kpc & Distance of 10 kpc  \\
         \hline
         $\rho_\text{gas}/(10^6 \text{M}_\odot \text{pc}^{-3})$ & $0.13 \pm 0.07$ & $0.2 \pm 0.1$\\

         $\rho_\text{ecl}/(10^6 \text{M}_\odot \text{pc}^{-3})$ & $0.03 \pm 0.02$ & $0.06 \pm 0.04$ \\
         
         $M_{\text{ecl}}/\text{M}_\odot$ & $180 \pm 8258$ & $1230 \pm 2636$ \\
         
         $r_\text{h}/\text{pc}$ & $0.082 \pm 0.005$ & $0.132 \pm 0.008$
    \end{tabular}
\end{table}

\begin{table}[h!]
    \caption{Results for the SC with an age of $1\,\mathrm{Myr}$.}
    \label{tab:sub_cluster_results_1}
    \centering
    \begin{tabular}{c c c}
    \hline \hline
         parameter & Distance of 2.5 kpc & Distance of 10 kpc  \\
         \hline
         $\rho_\text{gas}/(10^6 \text{M}_\odot \text{pc}^{-3})$ & $0.02 \pm 0.04$ & $0.04 \pm 0.07$\\

         $\rho_\text{ecl}/(10^6 \text{M}_\odot \text{pc}^{-3})$ & $0.01 \pm 0.01$ & $0.01 \pm 0.02$\\
         
         $M_{\text{ecl}}/\text{M}_\odot$ & $37 \pm 1619$ & $309 \pm 910$ \\
         
         $r_\text{h}/\text{pc}$ & $0.085 \pm 0.009$ & $0.15 \pm 0.02$
    \end{tabular}
\end{table}

\begin{figure}[H]
    \resizebox{\hsize}{!}{\includegraphics{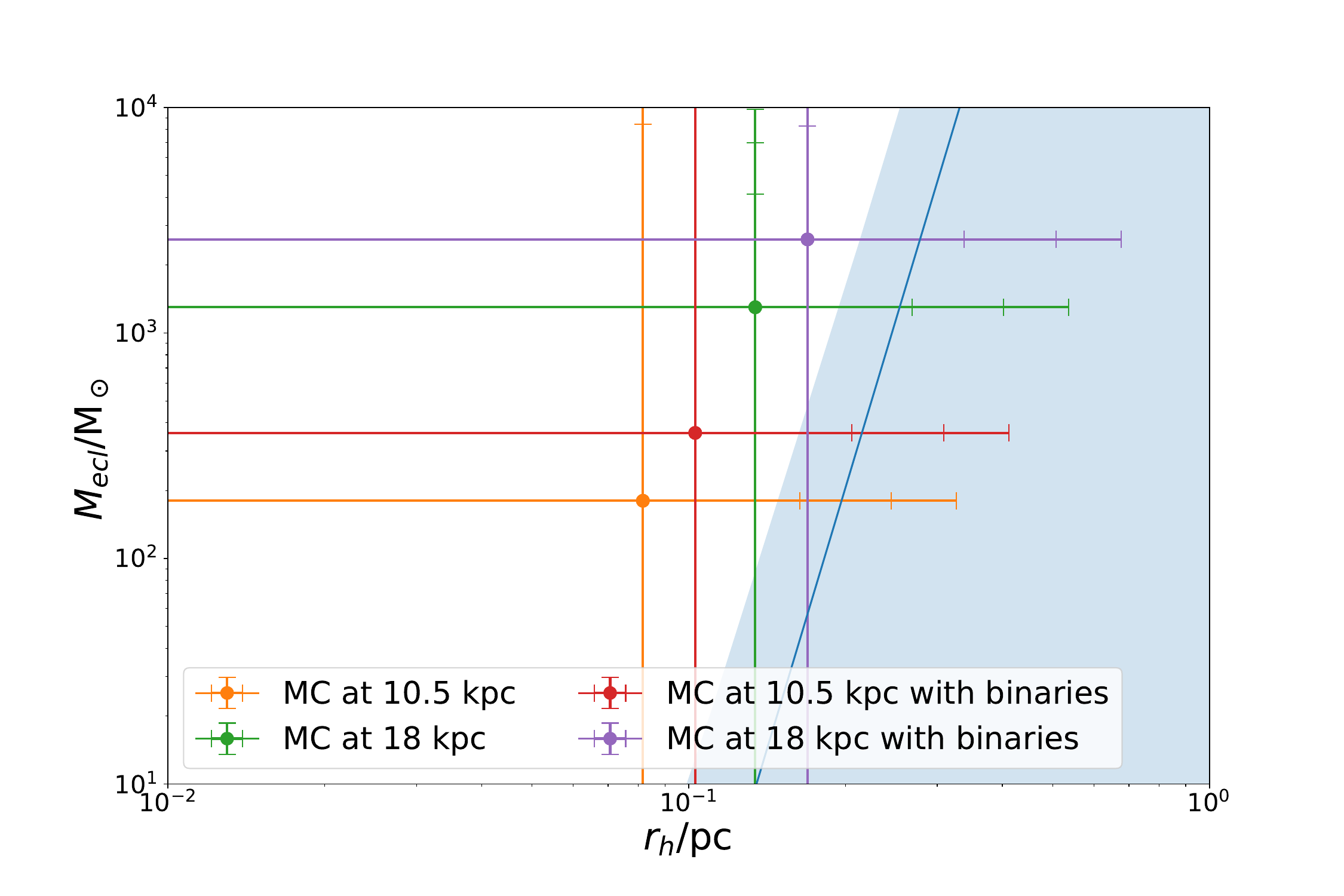}}
    \caption{As Fig.~\ref{fig:M_r_binary_MC}, but for the MC with an age of 1 Myr.}
    \label{fig:M_r_binary_MC_1}
\end{figure}

\begin{figure}[H]
    \resizebox{\hsize}{!}{\includegraphics{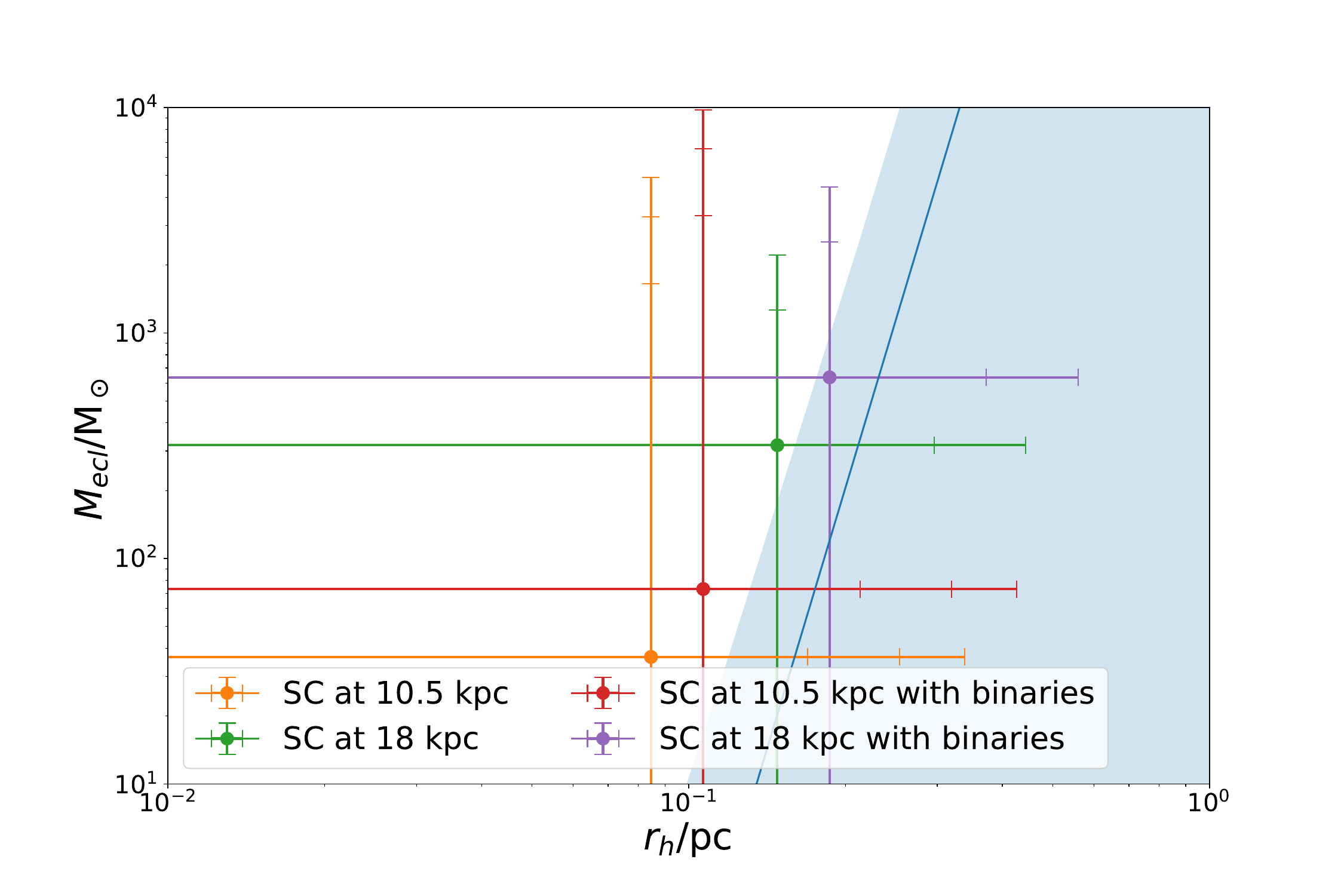}}
    \caption{As Fig.~\ref{fig:M_r_binary_SC}, but for the SC with an age of 1 Myr.}
    \label{fig:M_r_binary_SC_1}
\end{figure}

\subsection{Age of 5 Myr}

\begin{table}[h!]
    \caption{Best-fit values for the MC with an age of 5 Myr (\citealp{Yasui}).}
    \label{tab:main_cluster_values_5}
    \centering
    \begin{tabular}{c c c}
    \hline \hline 
     parameter & Distance of 2.5 kpc & Distance of 10 kpc \\
     \hline
    $\log_{10} m_1/\text{M}_\odot$ & $-1.2$ & $-0.3$ \\
    $\log_{10} m_2/\text{M}_\odot$  & $-0.9^{+0.1}_{-0.2}$ & $-0.2 \pm 0.1$\\
    $\alpha_1$  & 0.0 & 0.2\\
    $\alpha_2$ & $1.4$ & 0.9\\
    $\alpha_3$ & $1.0 \pm 0.1$ & $2.4 \pm 0.1$ \\
    \end{tabular}
\end{table}

\begin{table}[h!]
    \caption{Best-fit values for the SC with an age of 5 Myr (\citealp{Yasui}).}
    \label{tab:sub_cluster_values_5}
    \centering
    \begin{tabular}{c c c}
    \hline \hline 
     parameter & Distance of 2.5 kpc & Distance of 10 kpc \\
     \hline
    $\log_{10} m_1/\text{M}_\odot$ & $-1.0$ & $-0.4$ \\
    $\log_{10} m_2/\text{M}_\odot$  & $-0.7 \pm 0.2$ & $0.1 \pm 0.2$\\
    $\alpha_1$  & $-0.8$ & $0.4$\\
    $\alpha_2$ & $1.4$ & $1.1$\\
    $\alpha_3$ & $2.0 \pm 0.3$ & $2.7 \pm 0.3$ \\
    \end{tabular}
\end{table}

\begin{figure}[h!]
    \resizebox{\hsize}{!}{\includegraphics{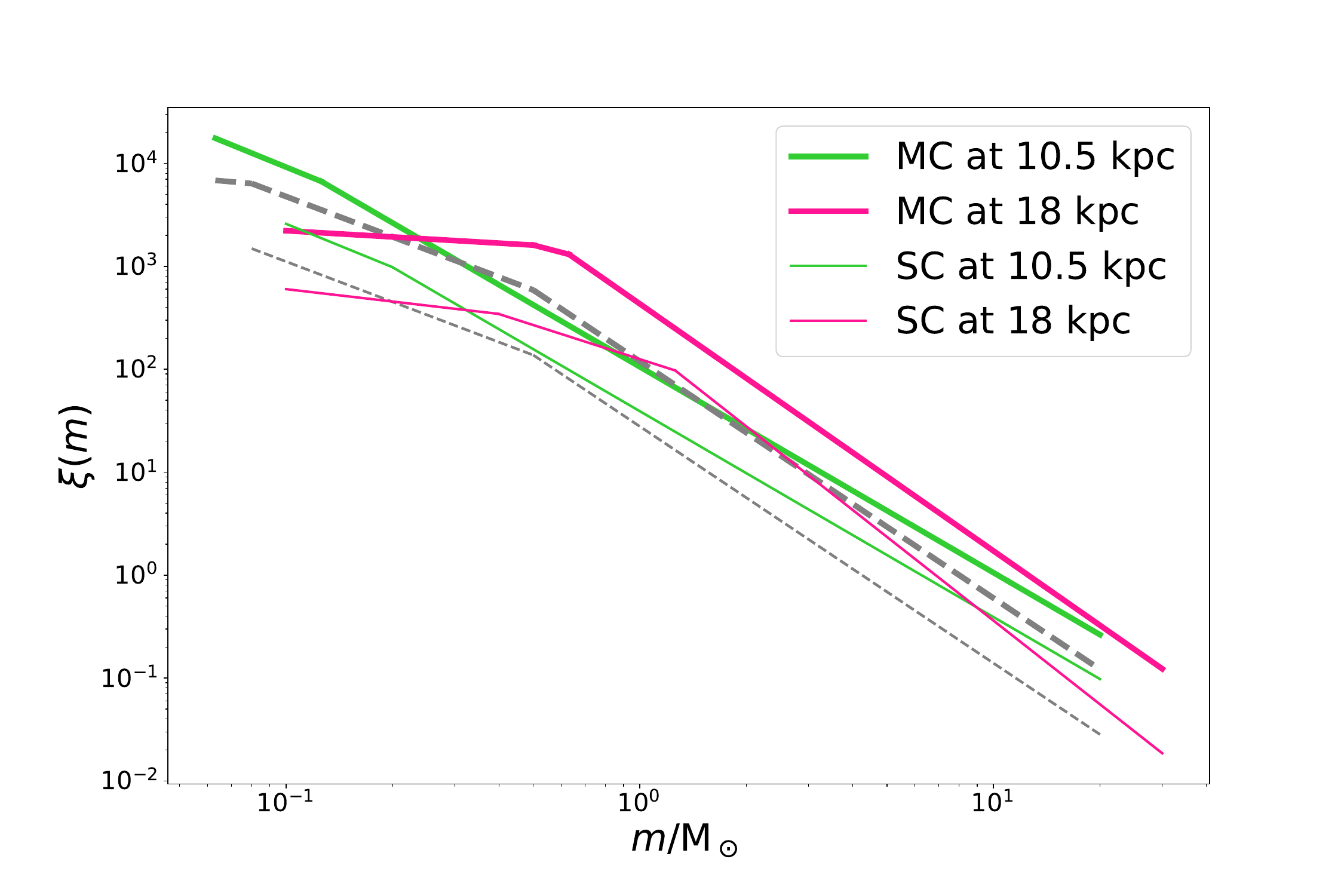}}
    \caption{As Fig.~\ref{fig:IMF}, but for an age of 5 Myr.}
    \label{fig:IMF_5}
\end{figure}

\begin{figure}[H]
    \resizebox{\hsize}{!}{\includegraphics{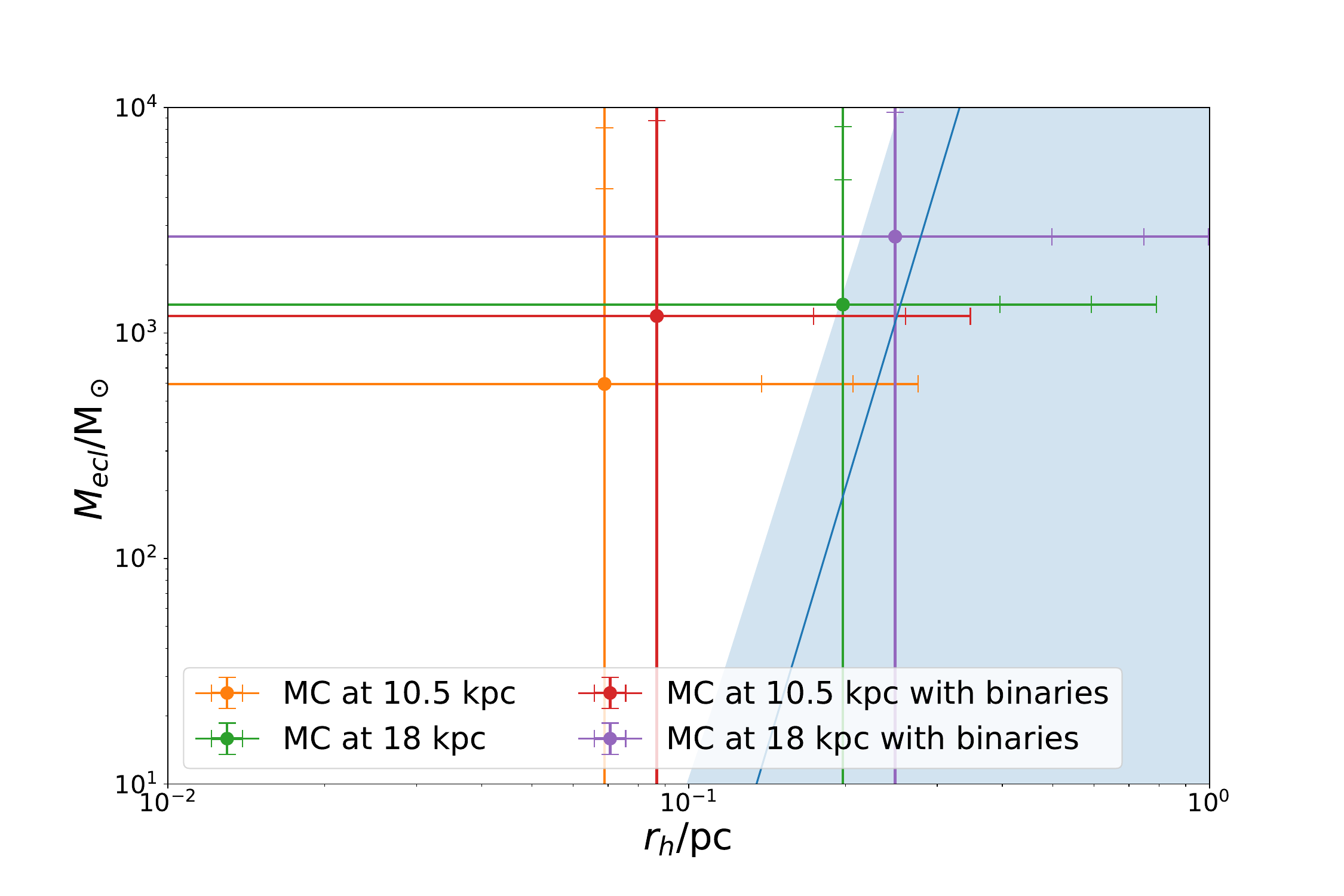}}
    \caption{As Fig.~\ref{fig:M_r_binary_SC}, but for the MC with an age of 5 Myr.}
    \label{fig:M_r_binary_MC_5}
\end{figure}

\begin{figure}[H]
    \resizebox{\hsize}{!}{\includegraphics{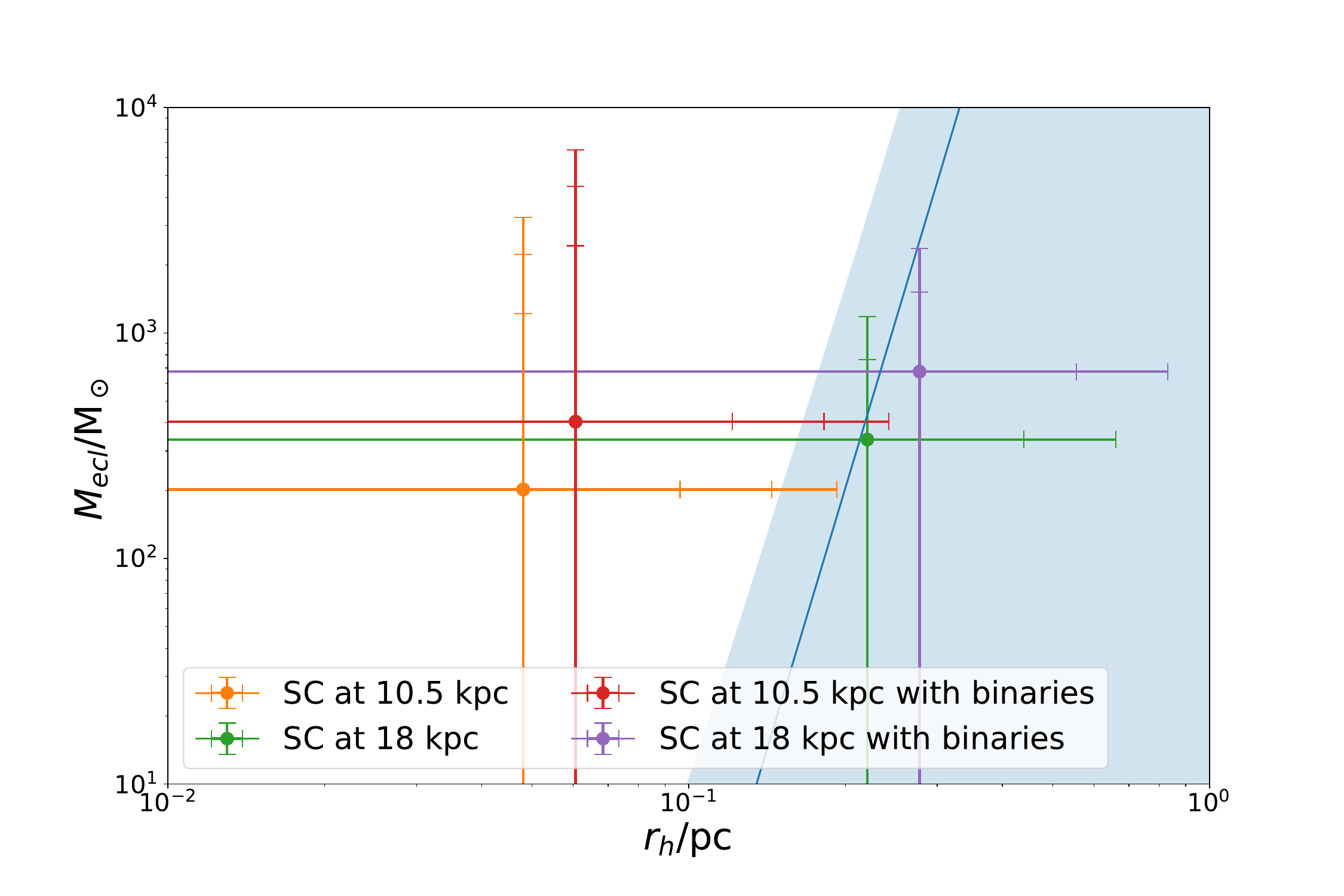}}
    \caption{As Fig.~\ref{fig:M_r_binary_SC}, but for the SC with an age of 5 Myr.}
    \label{fig:M_r_binary_SC_5}
\end{figure}

\begin{table}[h!]
    \caption{Results for the MC with an age of $5\,\mathrm{Myr}$.}
    \label{tab:main_cluster_results_5}
    \centering
    \begin{tabular}{c c c}
    \hline \hline
         parameter & Distance of 2.5 kpc & Distance of 10 kpc  \\
         \hline
         $\rho_\text{gas}/(10^6 \text{M}_\odot \text{pc}^{-3})$ & $0.7 \pm 0.4$ & $0.07 \pm 0.04$\\

         $\rho_\text{ecl}/(10^6 \text{M}_\odot \text{pc}^{-3})$ & $0.2 \pm 0.1$ & $0.02 \pm 0.01$ \\
         
         $M_{\text{ecl}}/\text{M}_\odot$ & $595 \pm 3773$ & $1291 \pm 3280$ \\
         
         $r_\text{h}/\text{pc}$ & $0.069 \pm 0.004$ & $0.20 \pm 0.01$
    \end{tabular}
\end{table}

\begin{table}[h!]
    \caption{Results for the SC with an age of $5\,\mathrm{Myr}$.}
    \label{tab:sub_cluster_results_5}
    \centering
    \begin{tabular}{c c c}
    \hline \hline
         parameter & Distance of 2.5 kpc & Distance of 10 kpc  \\
         \hline
         $\rho_\text{gas}/(10^6 \text{M}_\odot \text{pc}^{-3})$ & $0.7 \pm 1.2$ & $0.01 \pm 0.02$\\

         $\rho_\text{ecl}/(10^6 \text{M}_\odot \text{pc}^{-3})$ & $0.2 \pm 0.4$ & $0.004 \pm 0.006$\\
         
         $M_{\text{ecl}}/\text{M}_\odot$ & $202 \pm 1016$ & $330 \pm 407$ \\
         
         $r_\text{h}/\text{pc}$ & $0.048 \pm 0.005$ & $0.22 \pm 0.02$
    \end{tabular}
\end{table}

\subsection{Age of 10 Myr}

\begin{table}[H]
    \caption{Best-fit values for the MC with an age of 10 Myr (\citealp{Yasui}).}
    \label{tab:main_cluster_values_10}
    \centering
    \begin{tabular}{c c c}
    \hline \hline 
     parameter & Distance of 2.5 kpc & Distance of 10 kpc \\
     \hline
    $\log_{10} m_1/\text{M}_\odot$ & $-1.0$ & $-0.1$ \\
    $\log_{10} m_2/\text{M}_\odot$  & $-0.4 \pm 0.1$ & $0.0 \pm 0.1$\\
    $\alpha_1$  & $1.2$ & 0.0\\
    $\alpha_2$ & $0.7$ & $1.1$\\
    $\alpha_3$ & $3.0 \pm 0.1$ & $3.0 \pm 0.1$ \\
    \end{tabular}
\end{table}

\begin{table}[h!]
    \caption{Best-fit values for the SC with an age of 10 Myr (\citealp{Yasui}).}
    \label{tab:sub_cluster_values_10}
    \centering
    \begin{tabular}{c c c}
    \hline \hline 
     parameter & Distance of 2.5 kpc & Distance of 10 kpc \\
     \hline
    $\log_{10} m_1/\text{M}_\odot$ & $-1.0$ & $-0.1$ \\
    $\log_{10} m_2/\text{M}_\odot$  & $-0.3 \pm 0.2$ & $0.0 \pm 0.2$\\
    $\alpha_1$  & $-0.8$ & $-0.2$\\
    $\alpha_2$ & $1.4$ & $1.2$\\
    $\alpha_3$ & $2.5 \pm 0.3$ & $2.5 \pm 0.3$ \\
    \end{tabular}
\end{table}

\begin{figure}[h!]
    \resizebox{\hsize}{!}{\includegraphics{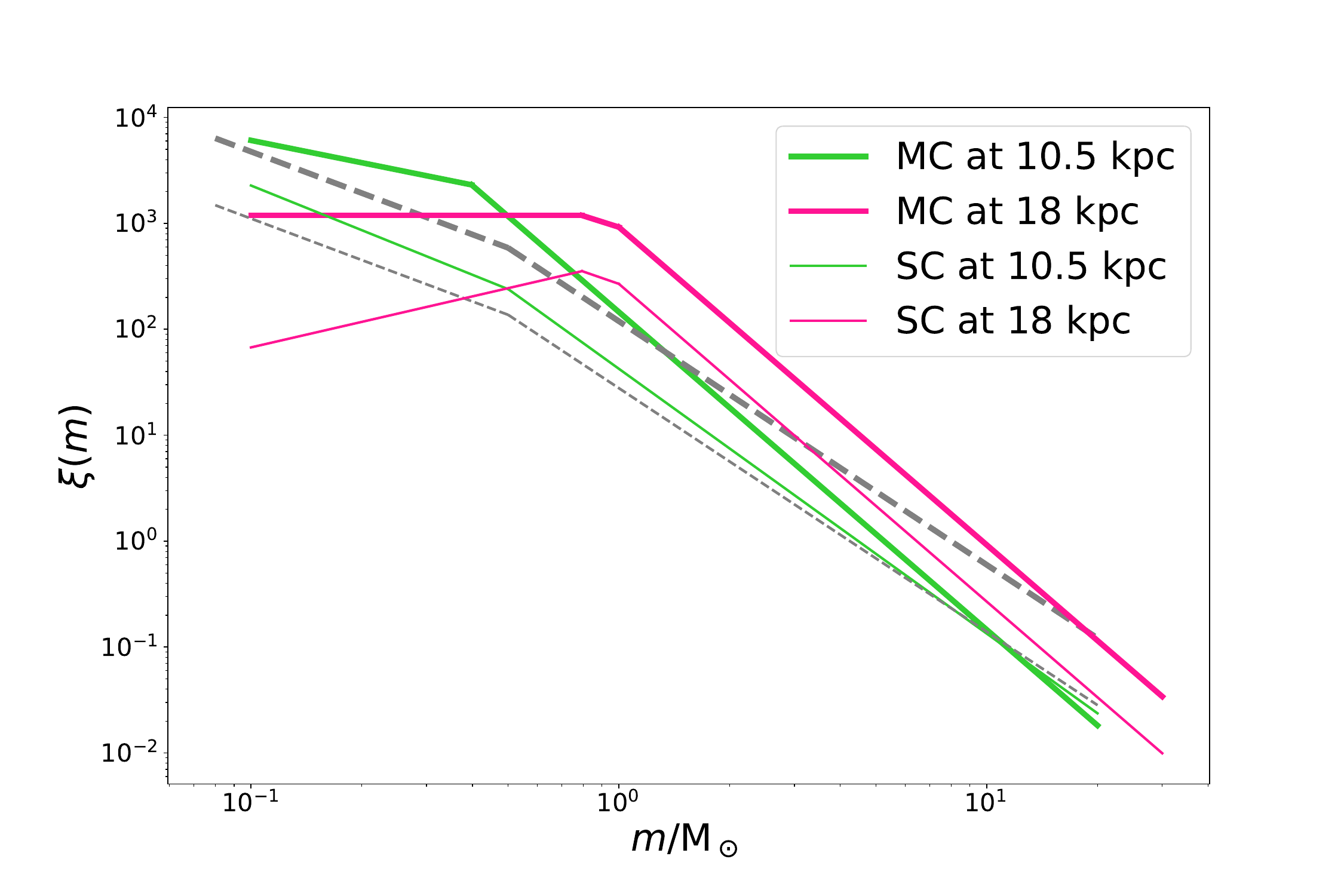}}
    \caption{As Fig.~\ref{fig:IMF}, but for an age of 10 Myr.}
    \label{fig:IMF_10}
\end{figure}

\begin{table}[h!]
    \caption{Results for the MC with an age of $10\,\mathrm{Myr}$.}
    \label{tab:main_cluster_results_10}
    \centering
    \begin{tabular}{c c c}
    \hline \hline
         parameter & Distance of 2.5 kpc & Distance of 10 kpc  \\
         \hline
         $\rho_\text{gas}/(10^6 \text{M}_\odot \text{pc}^{-3})$ & $0.002 \pm 0.001$ & $0.002 \pm 0.001$\\

         $\rho_\text{ecl}/(10^6 \text{M}_\odot \text{pc}^{-3})$ & $0.0007 \pm 0.0004$ & $0.0007 \pm 0.0004$ \\
         
         $M_{\text{ecl}}/\text{M}_\odot$ & $594 \pm 2927$ & $1437 \pm 2348$ \\
         
         $r_\text{h}/\text{pc}$ & $0.46 \pm 0.03$ & $0.63 \pm 0.04$
    \end{tabular}
\end{table}

\begin{table}[h!]
    \caption{Results for the SC with an age of $10\,\mathrm{Myr}$.}
    \label{tab:sub_cluster_results_10}
    \centering
    \begin{tabular}{c c c}
    \hline \hline
         parameter & Distance of 2.5 kpc & Distance of 10 kpc  \\
         \hline
         $\rho_\text{gas}/(10^6 \text{M}_\odot \text{pc}^{-3})$ & $0.04 \pm 0.07$ & $0.002 \pm 0.004$\\

         $\rho_\text{ecl}/(10^6 \text{M}_\odot \text{pc}^{-3})$ & $0.01 \pm 0.02$ & $0.0007 \pm 0.001$\\
         
         $M_{\text{ecl}}/\text{M}_\odot$ & $163 \pm 375$ & $391 \pm 847$ \\
         
         $r_\text{h}/\text{pc}$ & $0.12 \pm 0.01$ & $0.41 \pm 0.04$
    \end{tabular}
\end{table}

\begin{figure}[H]
    \resizebox{\hsize}{!}{\includegraphics{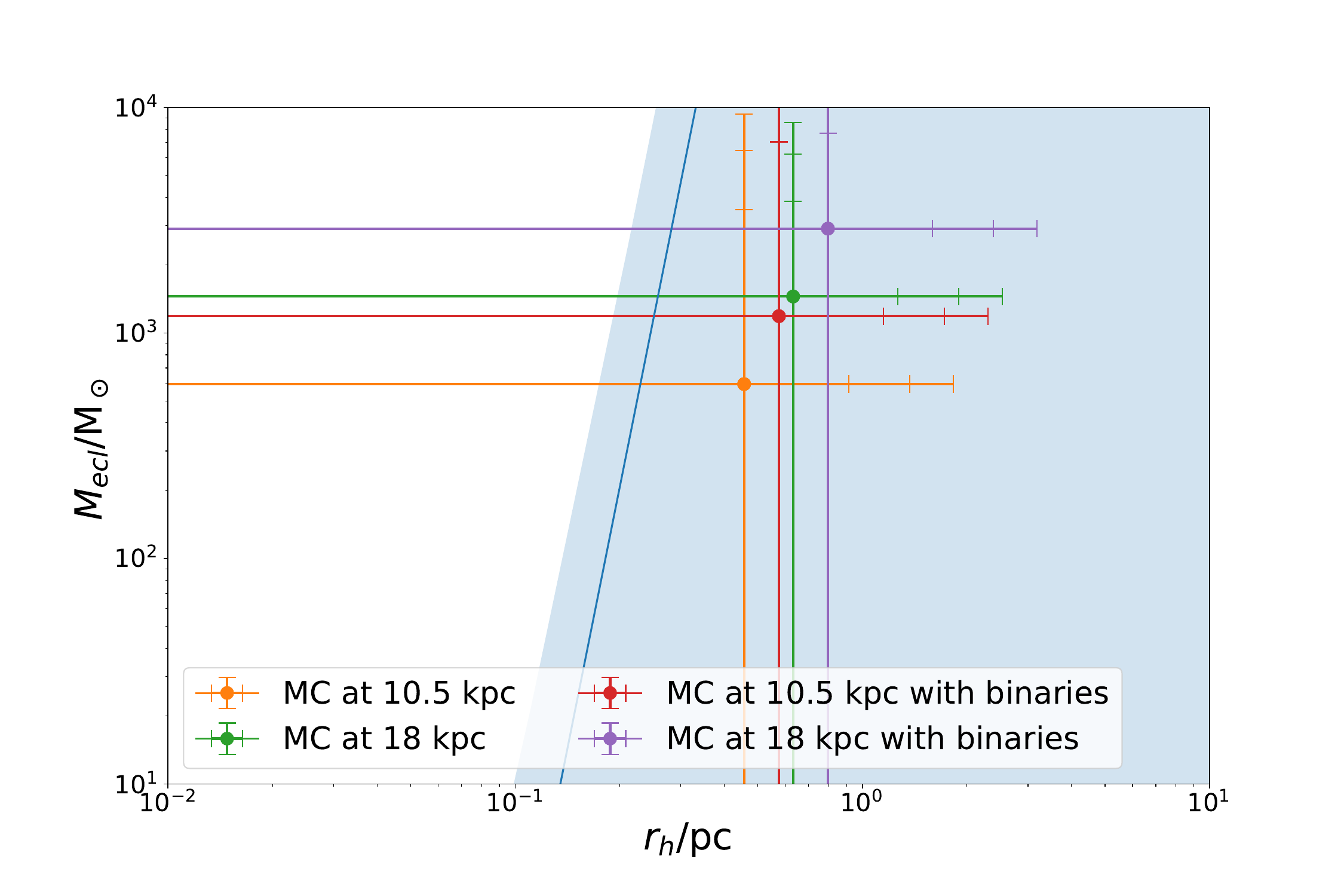}}
    \caption{As Fig.~\ref{fig:M_r_binary_MC}, but for the MC with an age of 10 Myr.}
    \label{fig:M_r_binary_MC_10}
\end{figure}

\begin{figure}[H]
    \resizebox{\hsize}{!}{\includegraphics{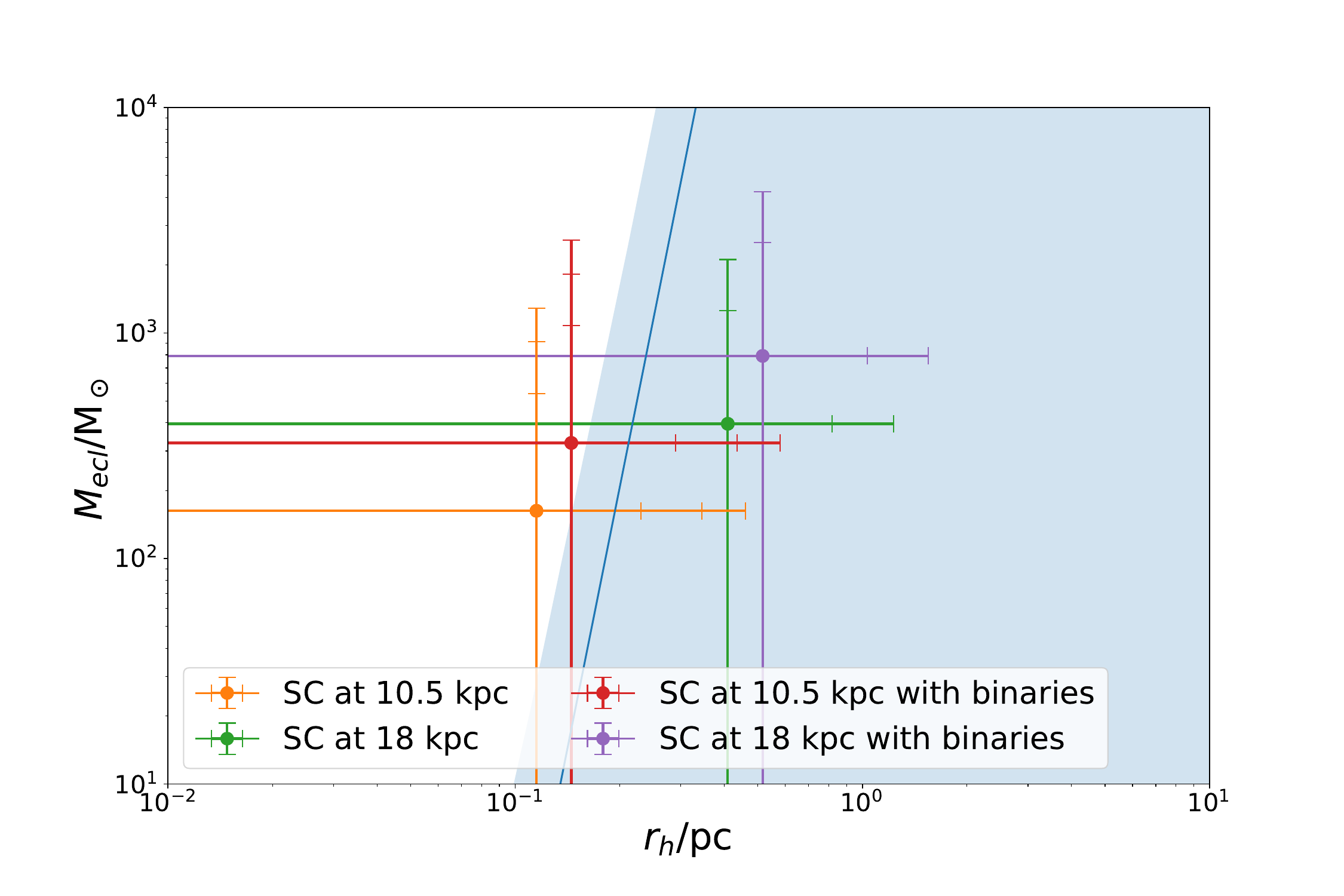}}
    \caption{As Fig.~\ref{fig:M_r_binary_SC}, but for the SC with an age of 10 Myr.}
    \label{fig:M_r_binary_SC_10}
\end{figure}

\end{appendix}

\end{document}